\documentclass[aps,prd,onecolumn,nofootinbib,superscriptaddress]{revtex4}
\pdfoutput=1
\usepackage{graphicx}
\usepackage{amsmath}
\usepackage{amsfonts}
\usepackage{amssymb,ulem}
\usepackage{color}%
\usepackage{dcolumn}

\usepackage{MnSymbol,wasysym}
\usepackage{braket}
\usepackage{eurosym}
\usepackage{calrsfs}
\usepackage{multirow}
\usepackage{float}
\usepackage[usenames,dvipsnames,svgnames]{xcolor}

\newcommand{\RNum}[1]{\uppercase\expandafter{\romannumeral #1\relax}}
\usepackage[colorlinks=true,linkcolor=blue,urlcolor=blue,filecolor=black,citecolor=red,pdfstartview=FitV,pdftitle={},pdfsubject={},pdfkeywords={},pdfpagemode=None,bookmarksopen=true]{hyperref}
\usepackage[title]{appendix}
\makeatletter
\newcommand*{\rom}[1]{\expandafter\@slowromancap\romannumeral #1@}
\makeatother

\begin{document}
\baselineskip=0.5 cm

\title{Perturbations of massless external fields in Horndeski hairy black hole}

\author{Zhen-Hao Yang}
\email{yangzhenhao$_$yzu@163.com}
\affiliation{Center for Gravitation and Cosmology, College of Physical Science and Technology, Yangzhou University, Yangzhou, 225009, China}

\author{Yun-He Lei}
\email{leiyunhe2022@163.com}
\affiliation{Center for Gravitation and Cosmology, College of Physical Science and Technology, Yangzhou University, Yangzhou, 225009, China}

\author{Xiao-Mei Kuang}
\email{xmeikuang@yzu.edu.cn (corresponding author)}
\affiliation{Center for Gravitation and Cosmology, College of Physical Science and Technology, Yangzhou University, Yangzhou, 225009, China}

\author{Jian-Pin Wu}
\email{jianpinwu@yzu.edu.cn}
\affiliation{Center for Gravitation and Cosmology, College of Physical Science and Technology, Yangzhou University, Yangzhou, 225009, China}

\begin{abstract}
\baselineskip=0.5 cm
In this paper, we study the propagations of external fields in Horndeski theory, including  the scalar field, electromagnetic field and Dirac field. We extensively explore the quasinormal frequencies, time evolution, greybody factors and emission rates of those massless perturbing fields  by solving the corresponding   master equations in the Horndeski hairy black hole. With the use of both numerical and analytical methods, we disclose the competitive/promotional influences of the Horndeski hair, spin and quantum momentum number of the external fields on those phenomenal physics. Our results show that the Horndeski hairy black hole is stable under those perturbations. Moreover, a larger Horndeski hair could enhance the intensity of energy emission rate for Hawking radiation of various particles, indicating that comparing to the Schwarzschild black hole, the Horndeski hariy black hole could have longer or shorter lifetime depending on the sign of the Horndeski hair.
\end{abstract}

\maketitle
\tableofcontents
\newpage
\section{Introduction}
Recent observational progress on gravitational waves \cite{LIGOScientific:2016aoc,LIGOScientific:2018mvr,LIGOScientific:2020aai} and black hole shadow \cite{EventHorizonTelescope:2019dse,EventHorizonTelescope:2022xnr} further demonstrates the great success of Einstein's general relativity (GR). Yet it is unabated that GR should be generalized, and in the generalized theories extra
fields or higher curvature terms are always involved in the action \cite{Nojiri:2006ri,Clifton:2011jh,Berti:2015itd}. Numerous modified gravitational theories were proposed, which indeed provide a richer framework and significantly help us further understand GR as well as our Universe. Among them, the scalar-tensor theories, which contain a scalar field $\phi$ as well as a metric tensor $g_{\mu\nu}$,  are known as the simplest nontrivial extensions of GR \cite{Damour:1992we}. One of the most famous four-dimensional scalar-tensor theory is the Horndeski gravity proposed in 1974 \cite{Horndeski:1974wa},  which contains higher derivatives of $\phi$ and $g_{\mu\nu}$ and is free of Ostrogradski instabilities because it possesses at most second-order differential field equations.
Various observational constraints or bounds on Horndeski theories have been explored in \cite{Bellini:2015xja,Bhattacharya:2016naa,Kreisch:2017uet,Hou:2017cjy,SpurioMancini:2019rxy,Allahyari:2020jkn}.

Horndeski gravity attracts lots of attention in the cosmological and astrophysical communities because it has significant consequences  in describing the accelerated expansion and other interesting features, please see \cite{Kobayashi:2019hrl} for review. Moreover,  Horndeski theory is important to test the no-hair theorem, because  it has  diffeomorphism invariance and second-order field equations, which are similar to  GR. In fact,  hairy black holes in Horndeski gravity have been widely constructed and analyzed, including the radially
dependent scalar field \cite{Rinaldi:2012vy,Cisterna:2014nua,Feng:2015oea,Sotiriou:2013qea, Miao:2016aol,Kuang:2016edj,Babichev:2016rlq,Benkel:2016rlz,Filios:2018xvy,Cisterna:2018hzf,Giusti:2021sku} and the time-dependent scalar field \cite{Babichev:2013cya,Babichev:2017lmw,BenAchour:2018dap,Takahashi:2019oxz,Minamitsuji:2019shy,Arkani-Hamed:2003juy}. However,  the hairy solution with scalar hair in linear time dependence was found to be unstable, and so this type of hairy solution was ruled out in Horndeski gravity \cite{Khoury:2020aya}.  Later in \cite{Hui:2012qt} ,  the no-hair theorem was  demonstrated not be hold when a Galileon field is coupled to gravity, but the static spherical black hole only  admits trivial Galileon profiles.  Then, inspired by \cite{Hui:2012qt}, the authors of \cite{Babichev:2017guv} further examined the no-hair theorem in Horndeski theories and beyond. They demonstrated that shift-symmetric Horndeski theory and beyond allow for static and asymptotically flat black holes with a nontrivial static scalar field,  and the action they considered is dubbed  quartic Horndeski gravity
\begin{eqnarray}\label{eq:action}
S=\int d^4x \sqrt{-g}\big[Q_2+Q_3\Box\phi+Q_4R+Q_{4,\chi}\left((\Box\phi)^2-(\nabla^\mu\nabla^\nu\phi)(\nabla_\mu\nabla_\nu\phi)\right)
+Q_5G_{\mu\nu}\nabla^\mu\nabla^\nu\phi\nonumber\\
-\frac{1}{6}Q_{5,\chi}\left((\Box\phi)^3-3(\Box\phi)(\nabla^\mu\nabla^\nu\phi)(\nabla_\mu\nabla_\nu\phi)
+2(\nabla_\mu\nabla_\nu\phi)(\nabla^\nu\nabla^\gamma\phi)(\nabla_\gamma\nabla^\mu\phi)\right)\big],
\end{eqnarray}
where $\chi=-\partial^\mu\phi\partial_\mu\phi/2$ is the canonical kinetic term, $Q_i~(i=2,3,4,5)$ are arbitrary functions of $\chi$ and $Q_{i,\chi} \equiv \partial Q_{i}/\partial \chi $, $R$ is the Ricci scalar and $G_{\mu\nu}$ is the Einstein tensor. In particular, very recently a static hairy black hole in a specific quartic Horndeski theory, saying that $Q_5$ in the above action vanishes, has been constructed in \cite{Bergliaffa:2021diw}
\begin{eqnarray}\label{eq:metric}
ds^2=-f(r)dt^2+\frac{dr^2}{f(r)}+r^2(d\theta^2+\sin^2\theta d\varphi^2)
~~~\mathrm{with}~~~f(r)=1-\frac{2M}{r}+\frac{Q}{r}\ln\left(\frac{r}{2M}\right).
\end{eqnarray}
Here, $M$ and $Q$ are the parameters related to the black hole mass and Horndeski hair.
The metric reduces to Schwarzschild case as $Q\to 0$, and it is asymptotically flat.
 From the metric \eqref{eq:metric}, it is not difficult to induce that for arbitrary $Q$, $r=0$ is an intrinsic singularity as the curvature scalar is singular, and $f(r)=0$ always admits a solution $r_+=2M$ which indicates a horizon at $r_+=2M$.  In addition, when $Q>0$, $r_+=2M$ is the unique root of $f(r)=0$ , so it has a unique horizon, i.e., the event horizon for the hairy black hole as for the Schwarzschild black hole. While for $-2M<Q<0$, $f(r)=0$ has two roots: one is $r_+=2M$ indicating the event horizon, and the other $r_-$ denotes the Cauchy horizon which is smaller than the event horizon. $r_-$ increases as $Q$ decreases, and finally approaches $r_+$ as $Q\to -2M$, meaning the extremal case.
This hairy black hole and its rotating counterpart attract plenty of attentions. Some theoretical and observational investigations have been carried out, for examples, the strong gravitational lensing \cite{Kumar:2021cyl}, thermodynamic and weak gravitational lensing \cite{Walia:2021emv,Atamurotov:2022slw}, shadow constraint from EHT observation \cite{Afrin:2021wlj},  superradiant energy extraction \cite{Jha:2022tdl} and photon rings in the black hole image \cite{Wang:2023vcv}.

However,  the propagation of external  field in the Horndeski hairy black hole is still missing. To fill this gap, here we shall  explore the quasinormal frequencies (QNFs), time evolution, greybody factors and emission rates  by analyzing the massless external  fields perturbations (including the scalar field, electromagnetic field and  Dirac field) around the Horndeski hairy black hole.

QNFs of a field perturbation on the black hole  are infinite discrete spectrum of complex frequencies,  of which the real part determines the oscillation timescale of the quasinormal modes (QNMs), while the complex part determines their exponential decaying timescale.
They dominate the signal of the gravitational waves at the ringdown stage  and are one of the most important characteristics of black hole geometry. The interest of QNMs in more fundamental physics can be referred to the reviews \cite{Nollert:1999ji,Berti:2009kk,Konoplya:2011qq}. Mathematically, QNFs depend solely on the basic three parameters of the black holes, i.e., the mass, charge, and angular momentum. However, if there are any additional parameters that describe the black hole, such as the hairy parameter in this framework, those parameters will also have prints on the QNMs spectrum.
Even though the propagations of external field  in a black hole background seems less related to the gravitational wave signals, they still might provide us with important insights about the properties of the Horndeski hairy black holes, such as their stability and the possible probe of the characterized parameters of black holes. This is our motivation to investigate the external  fields QNMs of the hairy black hole solution \eqref{eq:metric}. The goal is to study the influences of the hairy parameter $Q$ on the QNFs signature for the massless scalar field, electromagnetic field and  Dirac field perturbations, respectively. To this end, we will use both the WKB method and the matrix method to numerically obtain the QNFs, and also exhibit the time evolution of the perturbations in the time domain.

The other goal of this work is to study the impact  of Horndeski hair on the energy emission rate of the particles with spin $=0, 1$ and $1/2$, respectively,  and the greybody factors of their Hawking radiation from the Horndeski hairy black hole.  The grey-body factor  measures the modification of the pure black body spectrum and it is equal to the transmission probability of an outgoing wave radiated from the black hole event horizon to the asymptotic region \cite{Harmark:2007jy}.  It significantly describes information about the near-horizon structure of black holes \cite{Kanti:2002nr}. So, one can evaluate the energy emission rate of  Hawking radiation with the use of the greybody factor \cite{Hawking:1975vcx}. It is known that the Hawking radiation spectrum and its greybody factor are very sensitive to the modifications of GR. So they at least provide important sources of physical consequences for the modifications  in the formulation of the black hole. Thus, we could expect that the Horndeski hair will leave prints on the Hawking radiation spectrums as well as  the greybody factors. It is noted that the study of  the energy emission rates and greybody factors of the  particles with spin$=0, 1$ and $1/2$  requires one to solve the master equations of the scalar,   electromagnetic and  Dirac perturbing fields on the hairy black hole background. This process is similar to that we do when calculating the quasinormal modes, only with different boundary conditions:  the latter requires a purely outgoing wave at infinity and a purely ingoing wave at the event horizon,  while the former allows ingoing waves at infinity.

This paper is organized as follows. In section \ref{sec:eoms}, we show the master equations for the test massless scalar, electromagnetic and Dirac fields in the Horndeski hairy black hole, and analyze the properties of their effective potentials. In section \ref{sec:QNM}, we calculate the QNM frequencies of the perturbing fields with both the WKB method and matrix method, and then match the behaviors of the perturbations in the time domain. In section \ref{sec:Hawking Radiation}, by solving the corresponding master equations, we evaluate the greybody factor and the energy emission of Hawking radiation for various particles. The last section contributes to our conclusions and discussion.  Throughout the paper, we will set $c=\hbar=G=1$. Moreover, in a convenient way, we will  fix $M=1/2$ and denote $Q/M \to Q$ in all the computations.

\section{Master equations and effective potentials of the perturbing external fields}\label{sec:eoms}

In this section, we will show the master equations of various massless external fields, including the scalar field, electromagnetic field and Dirac field around the Horndeski hairy black hole. The influences of Horndeski hair $Q$ and angular quantum number $\ell$ on the corresponding effective potentials of the perturbations will be analysed.

\subsection{Scalar field perturbation}

The propagation of massless scalar field $\Phi$ in the Horndeski hairy black hole satisfies the Klein-Gordon equation
\begin{equation}\label{K--G Eq.}
\square\Phi=\frac{1}{\sqrt{-g}}\partial_\mu(g^{\mu\nu}\sqrt{-g}\partial_\nu\Phi)=0,
\end{equation}
where $g$ is the determinant of the black hole metric (\ref{eq:metric}). By taking the ansatz
\begin{equation}
\Phi(t,r,\theta,\varphi)=e^{-i\omega t+im\varphi}\frac{R(r)}{r}S(\theta),
\end{equation}
where $\omega$ is the frequency of scalar field perturbation and $m$ is the azimuthal number, we can separate \eqref{K--G Eq.} and obtain  the radial master equation
\begin{equation}\label{scalar radial Eq. in r}
f^2(r)\frac{d^2R}{dr^2}+f(r)f'(r)\frac{dR}{dr}+[\omega^2-V_{sc}(r)]R=0.
\end{equation}
Here, the prime denotes a derivative w.r.t. $r$, and the effective potential is
\begin{equation}\label{eq:Vsc}
V_{sc}(r)=\left[1-\frac{2}{r}\left(M-\frac{Q}{2}\ln\frac{r}{2M}\right)\right]\left[\frac{\ell(\ell+1)}{r^2}+\frac{1}{r^3}\left(Q+2M-Q\ln\frac{r}{2M}\right)\right]
\end{equation}
where $\ell=0,1,2,\cdots$ is the angular quantum number. Under the tortoise coordinate
\begin{equation}\label{eq:rstar}
r_*=\int\frac{dr}{f(r)},
\end{equation}
the equation \eqref{scalar radial Eq. in r} can be written into Schrodinger--like form
\begin{equation}\label{eq:scalar master eq}
\frac{d^2R}{dr_*^2}+[\omega^2-V_{sc}(r)]R=0.
\end{equation}

\begin{figure*}[htbp]
\includegraphics[height=4cm]{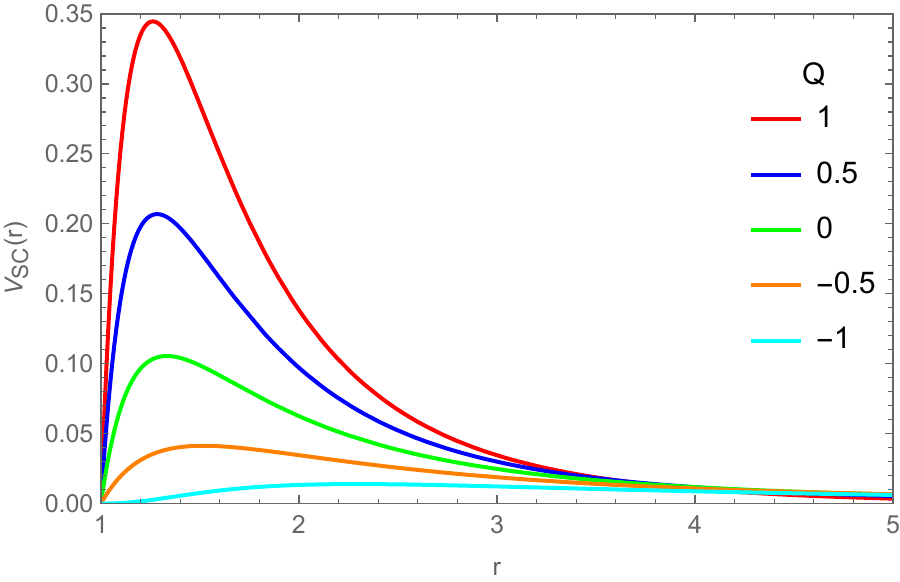}\hspace{0.5cm}
\includegraphics[height=4cm]{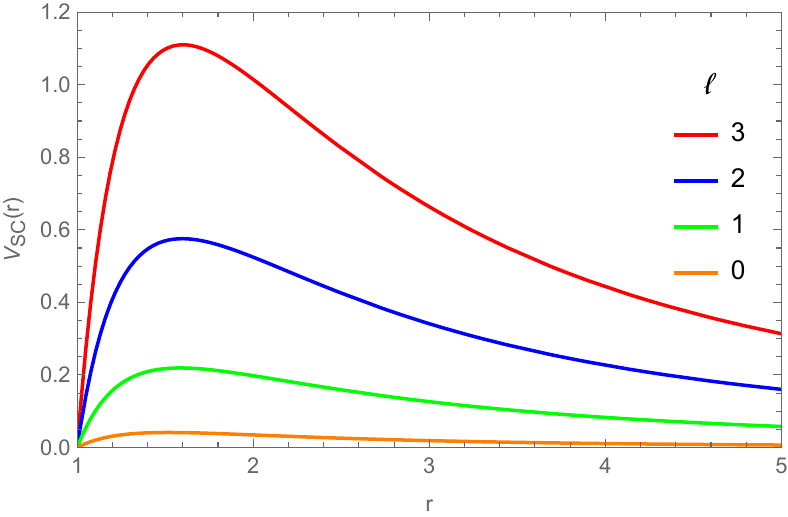}
\caption{\label{Fig:Vsc}The effective potential $V_{sc}(r)$ for the massless scalar field perturbation. In the left plot we fix $\ell=0$ and tune $Q$, while we fix $Q=-0.5$ and tune $\ell$ in the right plot, respectively.}
\end{figure*}

The behavior of effective potential $V_{sc}(r)$ with various samples of $Q$ and $\ell$ are shown in FIG.\ref{Fig:Vsc}. It can be obviously seen that, for each case with different $Q$ and $\ell$, the potential functions are always positive outside the event horizon. The absence of a negative potential well may give a hint that the black hole could remain stable under the massless scalar field perturbation. These plots also show that the effective potential always has a barrier near the horizon, which is enhanced by increasing the values  of $Q$ and  $\ell$.

\subsection{Electromagnetic perturbation}

\begin{figure*}[htbp]
\includegraphics[height=4cm]{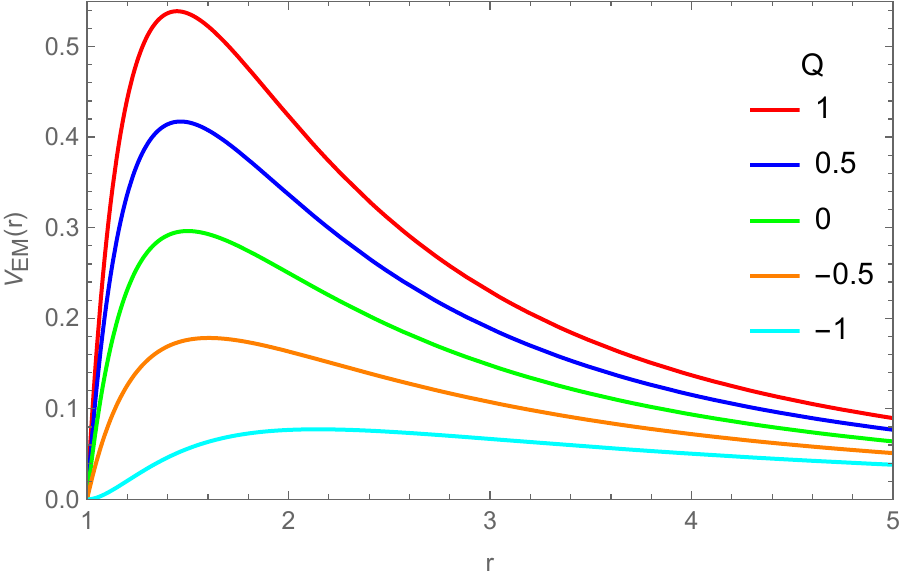}\hspace{0.5cm}
\includegraphics[height=4cm]{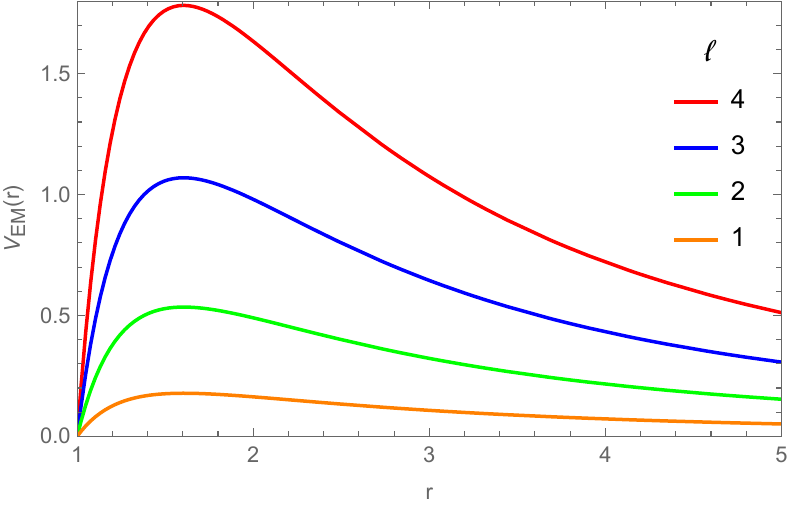}
\caption{\label{fig: V_EM}The effective potential of electromagnetic perturbation $V_{EM}(r)$. We fix $\ell=1$ and tune $Q$  in the left plot, while we fix $Q=-0.5$ and tune $\ell$ in the  right  plot, respectively.}
\end{figure*}

The propagation of electromagnetic field in the Horndeski hairy black hole background satisfies the Maxwell equation
\begin{eqnarray}\label{Maxwell Eq.}
\nabla_\nu F^{\mu\nu}=\frac{1}{\sqrt{-g}}\partial_\nu(\sqrt{-g}F^{\mu\nu})=0,
\end{eqnarray}
where $F_{\mu\nu}=\partial_\mu A_\nu-\partial_\nu A_\mu$ is the field strength tensor, and $A_\mu$ is the vector potential. In order to separate the Maxwell equation, we take $A_\mu$ as the Regge--Wheeler--Zerilli decomposition \cite{Regge:1957td,Zerilli:1970se}
\begin{eqnarray}
\label{RW decomposition}
A_\mu=e^{-i\omega t}\sum_{\ell ,m}\left[\begin{pmatrix}
0\\
0\\
a^{\ell m}(r) \, \frac{1}{\sin\theta} \, \partial_\varphi Y_{\ell m}\\
-a^{\ell m}(r) \, \sin\theta \, \partial_\theta Y_{\ell m}
\end{pmatrix}
+\begin{pmatrix}
j^{\ell m}(r) \, Y_{\ell m}\\
h^{\ell m}(r) \, Y_{\ell m}\\
k^{\ell m}(r) \, \partial_\theta Y_{\ell m}\\
k^{\ell m}(r) \, \partial_\varphi Y_{\ell m}
\end{pmatrix}\right],
\end{eqnarray}
where $Y_{\ell m}=Y_{\ell m}(\theta,\varphi)$ are the scalar spherical harmonics with the angular quantum number, $\ell$, and azimuthal number,  $m$, respectively.
By substituting Eq.\eqref{RW decomposition} into Eq.\eqref{Maxwell Eq.}, we will obtain two decoupled radial equations which can be uniformed into the master equation
\begin{equation}\label{EM eq in r}
f^2(r)\frac{d^2\psi}{dr^2}+f(r)f'(r)\frac{d\psi}{dr}+[\omega^2-V_{EM}(r)]\psi=0,
\end{equation}
with
\begin{eqnarray}
\psi(r)=\left\{ \begin{aligned}
&a^{\ell m}(r) \qquad \qquad \qquad \qquad \text{for axial modes with odd parity $(-1)^{\ell+1}$,}\\
\frac{r^2}{\ell(\ell+1)}&\left[i\omega h^{\ell m}(r)+\frac{dj^{\ell m}(r)}{dr}\right] \quad \text{for polar modes with even parity $(-1)^{\ell}$ .}
\end{aligned} \right.
\end{eqnarray}
Again under the tortoise coordinate (\ref{eq:rstar}), the Schrodinger-like form of Eq.\eqref{EM eq in r} reads as
\begin{equation}
\label{eq:EM master eq}
\frac{d^2\psi}{dr_*^2}+[\omega^2-V_{EM}(r)]\psi=0.
\end{equation}
where the effective potential is
\begin{equation}\label{eq:Vem}
V_{EM}(r)=\left[1-\frac{2}{r}\left(M-\frac{Q}{2}\ln\frac{r}{2M}\right)\right]\frac{\ell(\ell+1)}{r^2}.
\end{equation}
The potential function for the electromagnetic field perturbation is depicted in FIG.\ref{fig: V_EM}  which shows that both larger  $Q$ and $\ell$ give higher potential barrier, similar to the case in scalar field perturbation.


\subsection{Dirac field perturbation}
The propagation of massless fermionic field in the Horndeski hairy black hole background is governed by
\begin{equation}
    \Gamma^a [ \partial_a + \frac{1}{4} (\omega_{\mu \nu})_a \Gamma^{\mu \nu} ] \Psi= 0~~~~
 \mathrm{with}~~ \Gamma^{\mu \nu} =\frac{1}{2}[\Gamma^\mu,\Gamma^\nu]~~\mathrm{and}~~
 (\omega_{\mu \nu})_a  =(e_\mu)_b \nabla_a (e_\nu)^b.
\end{equation}
In this equation, $(\omega_{\mu \nu})_a$ is the 1-form spin connections; $(e_\mu)^a$ is a
rigid tetrad defined by $(e^\mu)_a=\sqrt{g_{\mu \nu}}(dx^\mu)_a$ and its dual form is  $(e_\mu)^a=(e^\nu)_b g^{ab} \eta_{\mu\nu}$ with $\eta_{\mu\nu}$ the Minkowski metric; $\Gamma^a$ is the curved spacetime gamma matrices, which connects the flat spacetime gamma via $\Gamma^a=(e_\mu)^a \Gamma^{\mu}$.
After working out the spin connections of the metric \eqref{eq:metric}, we expand the Dirac equations as
\begin{equation}
    \frac{\Gamma^0}{\sqrt{f}} \frac{\partial\Psi}{\partial t} +
    \Gamma^{1} \sqrt{f} (\frac{\partial}{\partial r} +\frac{1}{r} + \frac{f'}{4f})\Psi+
    \frac{\Gamma^2}{r}(\frac{\partial}{\partial \theta}+\frac{1}{2}cot\,\theta)\Psi+
    \frac{\Gamma^3}{r \, sin\,\theta} \frac{\partial \Psi}{\partial \varphi}=0.
\end{equation}
To proceed, we choose the representation of the flat spacetime gamma matrices as  \cite{Cho:2003qe}
\begin{equation}
    \Gamma^0= \begin{pmatrix} -i & 0 \\ 0 & i \end{pmatrix}, \quad
    \Gamma^i= \begin{pmatrix} 0 & -i\sigma^i \\ i\sigma^i & 0 \end{pmatrix}, \quad
    i=1,2,3,
\end{equation}
where $\sigma^i$ are the Pauli matrices.
Then considering the Dirac field decomposition \cite{Cho:2003qe}
\begin{equation}
    \Psi^{(\pm)}(t,r,\theta,\varphi)= \frac{e^{-i\omega t}}{r f(r)} \begin{pmatrix} iG^{(\pm)}(r) \\ F^{(\pm)}(r) \end{pmatrix} \otimes  \begin{pmatrix} \phi^{(\pm)}_{jm}(\theta,\varphi) \\ \phi^{(\mp)}_{jm}(\theta,\varphi) \end{pmatrix},
\end{equation}
where the spinor angular harmonics are
\begin{equation}
\begin{split}
        \phi^{(+)}_{jm}=\begin{pmatrix} \sqrt{\frac{j+m}{2j}} Y^{m-1/2}_{l} \\ \sqrt{\frac{j-m}{2j}} Y^{m+1/2}_{l} \end{pmatrix} \qquad \text{for} \quad j=l+\frac{1}{2},\\
    \phi^{(-)}_{jm}=\begin{pmatrix} \sqrt{\frac{j+1+m}{2(j+1)}} Y^{m-1/2}_{l} \\ -\sqrt{\frac{j+1-m}{2(j+1)}} Y^{m+1/2}_{l} \end{pmatrix} \qquad \text{for} \quad j=l-\frac{1}{2},
    \end{split}
\end{equation}
we can obtain two radial master equations
\begin{eqnarray}
    r^2 f \, \partial_{r}(f \, \partial_{r}G^{{(\pm)}}) +(r^2\omega^2-\kappa^2_{\pm}f-\kappa_{\pm}f^{3/2}+\frac{1}{2}\kappa\sqrt{f}f')G^{(\pm)} =0, \\
    r^2 f \, \partial_{r}(f \, \partial_{r}F^{{(\pm)}}) +(r^2\omega^2-\kappa^2_{\pm}f+\kappa_{\pm}f^{3/2}-\frac{1}{2}\kappa\sqrt{f}f')F^{(\pm)} =0.
\end{eqnarray}
where $\kappa_{\pm}=\mp (j+\frac{1}{2})$ for $j=l \pm 1/2$.
The Schrodinger-like equations under the tortoise coordinate take the forms
\begin{eqnarray}
\label{eq:Dirac master eq}
    \frac{d^2F^{(\pm)}}{dr^2_*}+[\omega^2-V^{\rom{1}}_{Dirac}]F^{(\pm)}=0, \\
     \frac{d^2G^{(\pm)}}{dr^2_*}+[\omega^2-V^{\rom{2}}_{Dirac}]G^{(\pm)}=0,
\end{eqnarray}
where the effective potentials are
\begin{eqnarray} \label{eq:Vdirac}
V^{\rom{1}}_{Dirac}=\frac{\sqrt{f}|\kappa_{+}|}{r^2}(|\kappa_{+}|\sqrt{f}+\frac{r f'}{2}-f), \\
    V^{\rom{2}}_{Dirac}=\frac{\sqrt{f}|\kappa_{-}|}{r^2}(|\kappa_{-}|\sqrt{f}-\frac{r f'}{2}+f).
\end{eqnarray}

\begin{figure*}[htbp]
\includegraphics[height=4cm]{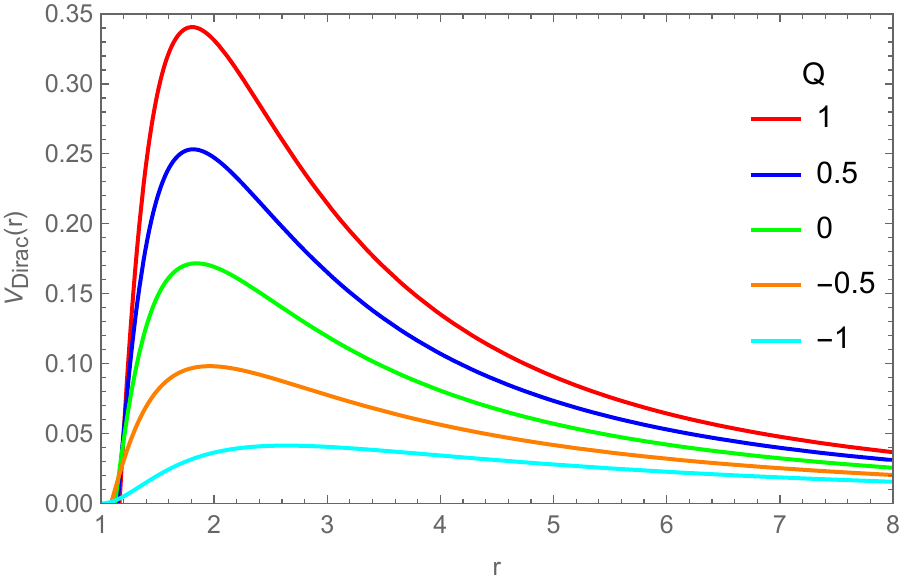}\hspace{0.5cm}
\includegraphics[height=4cm]{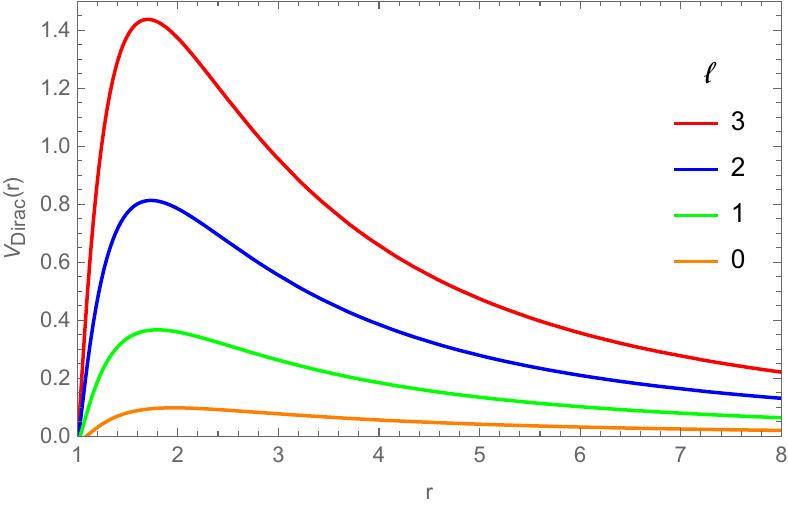}
\caption{\label{fig:V_Dirac}The effective potential of Dirac perturbation $V_{Dirac}^I(r)$. We fix $\ell=0$ and tune $Q$  in the left plot, while we fix $Q=-0.5$ and tune $\ell$ in the right   plot.}
\end{figure*}

It was addressed in \cite{Anderson:1991kx} that the behaviors of  $V_{Dirac}^I$ and $V_{Dirac}^{II}$ usually are qualitatively similar because they are super-symmetric partners derived from the same super potential, so one can choose one of them to proceed without any loss of generality.  Thus, in the following study, we will concentrate on the master equation \eqref{eq:Dirac master eq}  with $V_{Dirac}^I$,  which is plotted in FIG.\ref{fig:V_Dirac} for some references of $Q$ and $\ell$.

Comparing these three effective potentials in FIG.\ref{Fig:Vsc}-FIG.\ref{fig:V_Dirac} for various perturbing external fields,  we can extract the following properties. (i) The Horndeski hair has the same effect on the maximum value of various potentials, i.e., a larger $Q$ corresponds to a larger and narrower barrier, from which we expect the same influence on the QNFs of those perturbations. (ii) The effect of $\ell$ on the behavior of various potentials in Horndeski hairy black hole are similar to that in the Schwarzschild black hole \cite{Kokkotas:1999bd}. (iii) The barrier of the effective potential for perturbing field with higher spin seems larger and wider. We could expect that the features of effective potentials could be reflected in the QNFs.

\section{Quasi-normal mode frequencies of various perturbations}\label{sec:QNM}
In this section, we shall compute the quasi-normal frequencies of the
Horndeski hairy black hole under the massless scalar, EM and Dirac fields perturbations by solving the master equations \eqref{eq:scalar master eq}, \eqref{eq:EM master eq} and \eqref{eq:Dirac master eq} with the boundary conditions: ingoing wave ($\sim e^{-i\omega r_*}$) at the horizon and the outgoing wave ($\sim e^{i\omega r_*}$) at infinity. We will employ both the WKB method and matrix method  to guarantee that we focus on the fundamental mode with the node $n=0$, and also to pledge  the  credibility of our results. Moreover, in order to directly analyze the essence of QNFs in the propagation of various perturbations, we will also study the time evolution of the perturbation fields with the use of time domain integration.  Instead of the tedious details in the main part, we briefly review the main steps of WKB method, matrix method  and  domain integration method of our framework in appendixes \ref{sec:appendix1}-\ref{sec:appendix3}, because all the methods are widely used in the related studies. Especially, we use the Pad\text{\'e} series $P^{\Tilde{n}}_{\Tilde{m}}$ \cite{Matyjasek:2017psv} in WKB method to improve and reconstruct the WKB correction terms and transform it into a continued-fraction-like form. Here $\Tilde{n}$ and $\Tilde{m}$ are the order numbers of the numerator and denominator series (see \cite{Konoplya:2019hlu}).

\subsection{$Q-$ dependence}
\begin{figure*}[htbp]
\includegraphics[height=4cm]{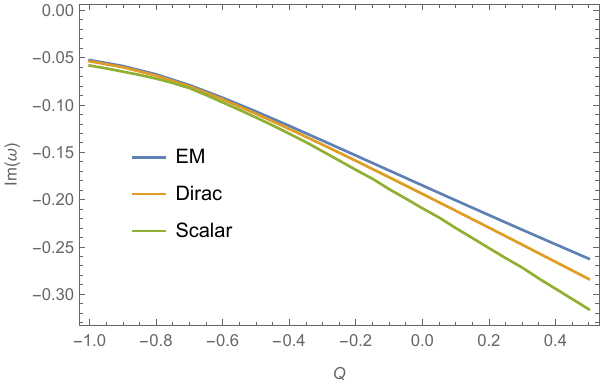}\hspace{0.5cm}
\includegraphics[height=4cm]{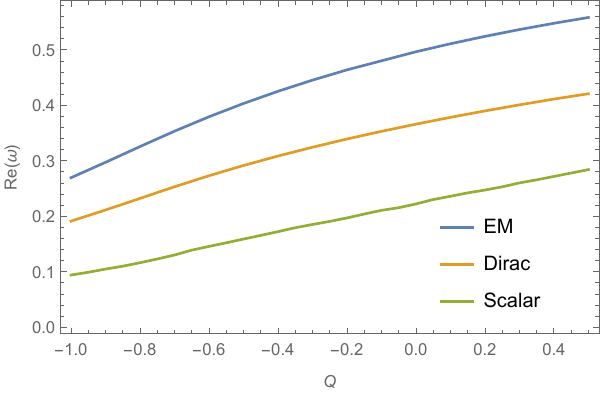}
\caption{\label{matterfieldsQNFl0}Quasi-normal frequencies  as a function of the hairy charge at the low-lying angular quantum number, i.e., $\ell=0$ for scalar $\&$ Dirac fields and $\ell=1$ for EM field. }
\end{figure*}

Firstly, we analyze the influence of the Horndeski hair parameter $Q$ on the QNM frequencies of the lowest $\ell$ modes in various external field perturbations.
\begin{table*}[h!]
	\center
	\begin{tabular}{|c|c|c|c|c|} \hline
		\null & \multicolumn{2}{c|}{ scalar field ($s=0$),\quad$\ell=0$} &\multicolumn{2}{c|}{relative error/\%}  \\ \hline
		$Q$ & Matrix Method  & WKB-Pad\text{\' e} & Re($\omega$)  & Im($\omega$)  \\ \hline
		0.5& 0.281229 - 0.318757 i & 0.284137 - 0.315827 i & -1.0234 & 0.9277\\ \hline
    0.2& 0.244945 - 0.252558 i & 0.247241 - 0.251195 i & -0.9286 & 0.5426\\ \hline
      0& 0.220902 - 0.209793 i & 0.222226 - 0.209131 i & -0.5958 & 0.3165\\ \hline
   -0.2& 0.196742 - 0.168829 i & 0.196977 - 0.168488 i & -0.1193 & 0.2024\\ \hline
   -0.5& 0.158948 - 0.113060 i & 0.158921 - 0.113149 i & 0.0170 & -0.0787\\ \hline
   -0.9& 0.104702 - 0.064905 i & 0.104821 - 0.064700 i & -0.1135 & 0.3168\\ \hline
   -1.0& 0.094461 - 0.058548 i & 0.094612 - 0.058225 i & -0.1596 & 0.5547\\ \hline\hline
\null & \multicolumn{2}{c|}{ Dirac field ($s=1/2$),\quad $\ell=0$ } &\multicolumn{2}{c|}{relative error/\%}  \\ \hline
		$Q$ & Matrix Method  & WKB-Pad\text{\' e} & Re($\omega$)  & Im($\omega$)  \\ \hline
  0.5& 0.420543 - 0.284416 i & 0.420877 - 0.283906 i & -0.0794 & 0.1796\\ \hline
  0.2& 0.389565 - 0.229705 i & 0.389876 - 0.229584 i & -0.0798 & 0.0527\\ \hline
  0  & 0.365925 - 0.193964 i & 0.366055 - 0.193908 i & -0.0355 & 0.0289\\ \hline
  -0.2& 0.339191 - 0.159007 i & 0.339260 - 0.158923 i & -0.0203 & 0.0529\\ \hline
  -0.5& 0.291568 - 0.109287 i & 0.291567 - 0.109216 i & 0.0003 & 0.0650\\ \hline
  -0.9& 0.211299 - 0.060102 i & 0.211320 - 0.060095 i & -0.0099 & 0.0116\\ \hline
  -1.0& 0.191377 - 0.053769 i & 0.191306 - 0.053776 i & 0.0371 & -0.0130\\ \hline\hline
\null & \multicolumn{2}{c|}{ electromagnetic field ($s=1$),\quad $\ell=1$} &\multicolumn{2}{c|}{relative error/\%}  \\ \hline
		$Q$ & Matrix Method  & WKB-Pad\text{\' e} & Re($\omega$)  & Im($\omega$)  \\ \hline
	   0.5& 0.558300 - 0.262397 i & 0.558302 - 0.262373 i & -0.0004 & 0.0091\\ \hline
   0.2& 0.524256 - 0.216346 i & 0.524253 - 0.216338 i & 0.0006  & 0.0037\\ \hline
   0  & 0.496527 - 0.184975 i & 0.496519 - 0.184966 i & 0.0016  & 0.0049\\ \hline
  -0.2& 0.463875 - 0.153360 i & 0.463859 - 0.153363 i & 0.0034  & -0.0020\\ \hline
  -0.5& 0.403289 - 0.106855 i & 0.403290 - 0.106854 i & -0.0002 & 0.0009\\ \hline
  -0.9& 0.297249 - 0.058808 i & 0.297253 - 0.058796 i & -0.0013 & 0.0204\\ \hline
  -1.0& 0.269519 - 0.052500 i & 0.269520 - 0.052497 i & -0.0004 & 0.0057\\ \hline
 \end{tabular}
  \caption{The fundamental ($n=0$) QNFs of lowest $\ell$-mode for various massless field perturbations obtained by WKB method and matrix method, and their relative errors.\label{Table:QNM girds varying q}}
\end{table*}
The results are listed in TABLE \ref{Table:QNM girds varying q} (also depicted in FIG. \ref{matterfieldsQNFl0}) in which we also calculate the relative error between two methods defined by
 \begin{eqnarray}
 \label{relative error}
     \Delta_k=\frac{\text{Matrix}(\omega_k)-\text{WKBP}(\omega_k)}{\text{WKBP}(\omega_k)}100\%,\qquad k=\text{Re}, \text{Im}.
 \end{eqnarray}

The QNFs obtained from the matrix and WKB-Pad\text{\'e} methods agree well with each other. For various perturbations, the imaginary part of QNFs, $Im(\omega)$, keeps increasing as $Q$ decreases, and it is always negative even in the extremal case with $Q=-1$. It means that the Horndeski hairy black hole is dynamically stable under those external fields perturbation with the lowest-lying $\ell$. Moreover, by comparing the QNFs for various perturbations, we find that for the field with larger spin, the $Im(\omega)$ is larger (see also the left plot of FIG. \ref{matterfieldsQNFl0} ). Thus, the perturbation field with a higher spin could live longer than the one with a lower spin because the  damping time $\tau_d$ for a wave field is related with the QNF by $\tau_d\sim 1/|-Im(\omega)|$. However,  the real part of QNFs, $Re(\omega)$, for all the perturbations decreases as $Q$ decreases, meaning that smaller $Q$ suppresses the oscillation of the perturbations. Similar to the imaginary part, $Re(\omega)$ is larger for a perturbing field with higher spin.  The effects of Horndeski hair and the spin of fields on the QNFs can be explained by their influences on  the corresponding effective potentials as we described in the previous section.
\begin{figure*}[htbp]
\includegraphics[height=3.2cm]{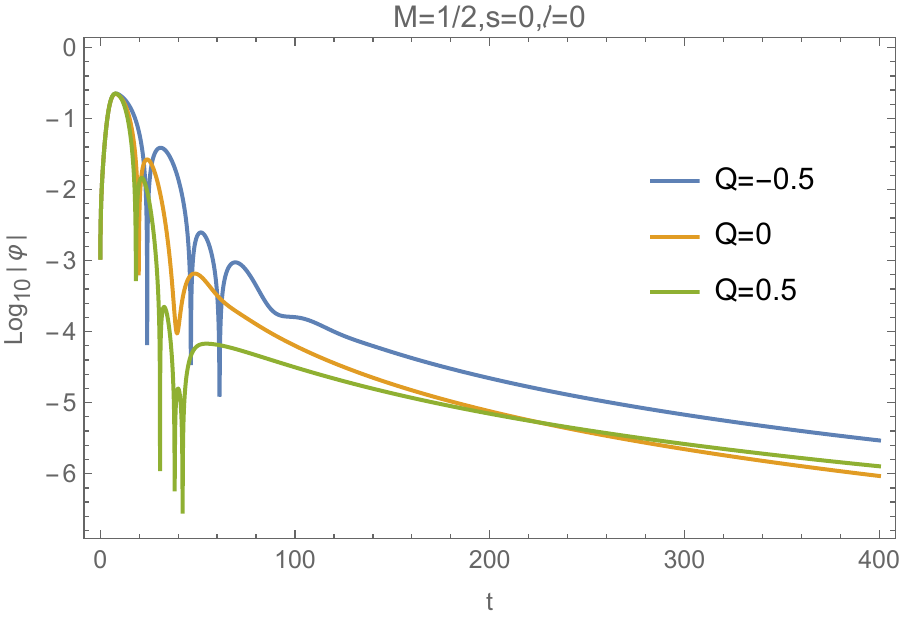}\hspace{0.5cm}
\includegraphics[height=3.2cm]{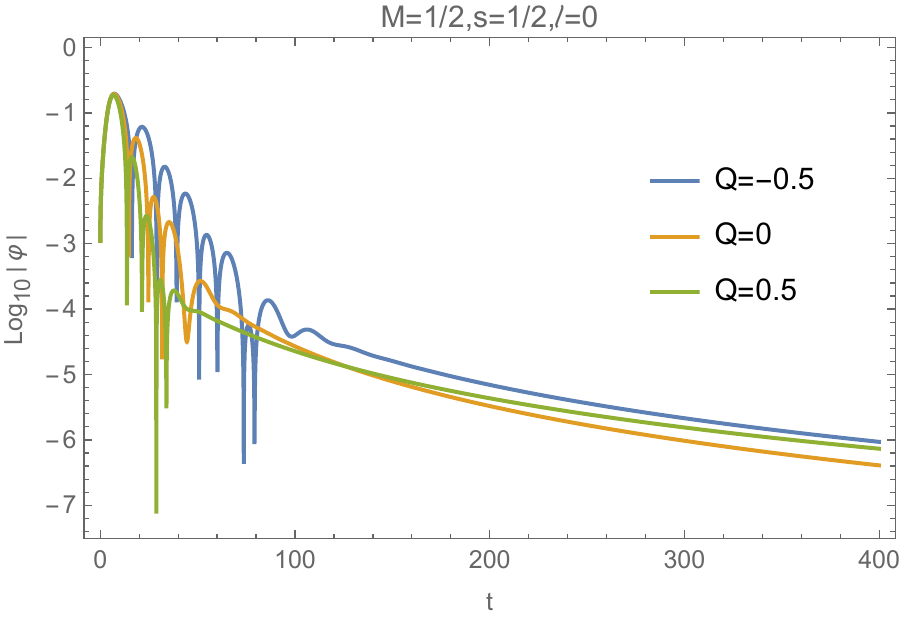}\hspace{0.5cm}
\includegraphics[height=3.2cm]{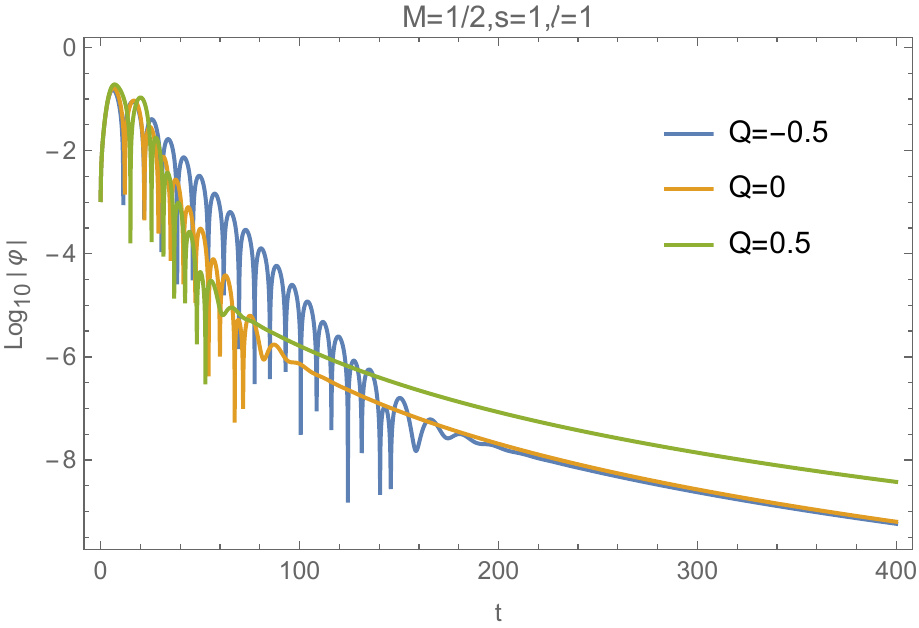}
\caption{\label{Timedomain of all matter fields0} Time evolution of the lowest-lying mode for the perturbing scalar field with spin $s=0$ (left), Dirac field with $s=1/2$ (middle) and EM field with $s=1$ (right), respectively. }
\end{figure*}

For the sake of intuitively understanding how the hairy charge $Q$ influences the evolution and waveform of various perturbations, we use the finite difference method to numerically integrate the wave-like equations \eqref{eq:scalar master eq}, \eqref{eq:EM master eq}, \eqref{eq:Dirac master eq} in the time domain. For more details, the readers can refer to the appendix \ref{sec:appendix3}. The results for the lowest-lying modes are shown in FIG.\ref{Timedomain of all matter fields0}. For all perturbations in each plot, we observe that smaller $Q$ makes  the ringing stage of perturbations waveform more lasting and sparser,  which corresponds to the larger $Im(\omega)$ and the smaller $Re(\omega)$. In addition, by a careful comparison,  we also find that the perturbation wave  with higher spin will perform a shorter and more intensive ringing stage, indicating smaller  $Im(\omega)$  and larger $Re(\omega)$. These observations in time domains agree well with the results we obtained in the frequency domain, also they explicitly show the time evolution process of various perturbing fields.

\subsection{$\ell-$ dependence}
\begin{figure*}[htbp]
\includegraphics[height=4cm]{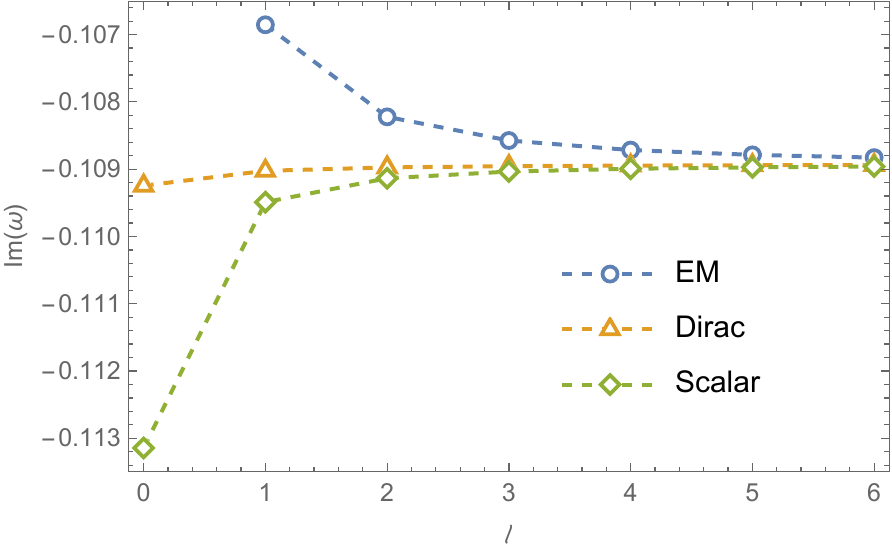}\hspace{0.5cm}
\includegraphics[height=4cm]{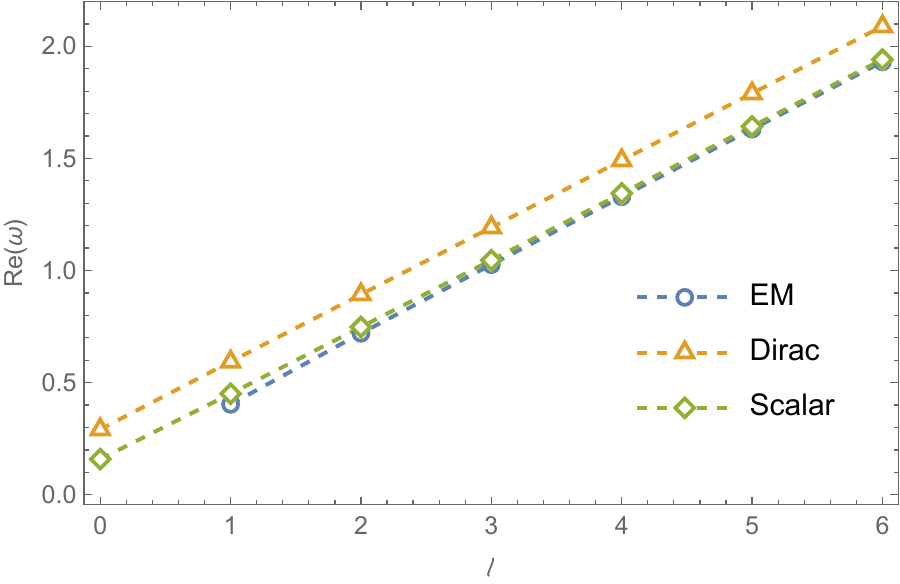}
\caption{\label{matterfieldsQNFl1} Quasi-normal frequencies as a function of the angular momentums for various perturbing fields.}
\end{figure*}

\begin{figure*}[htbp]
\includegraphics[height=3.2cm]{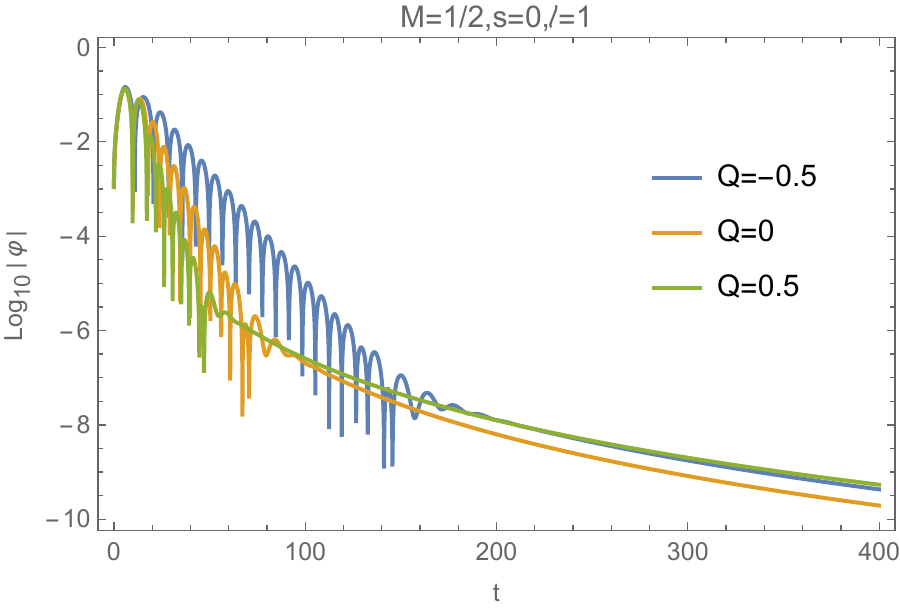}\hspace{0.5cm}
\includegraphics[height=3.2cm]{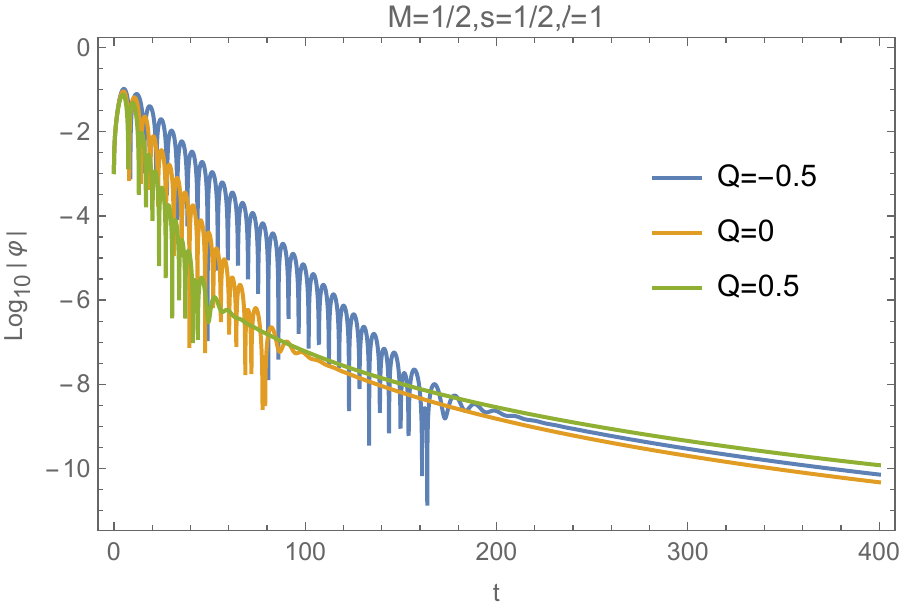}\hspace{0.5cm}
\includegraphics[height=3.2cm]{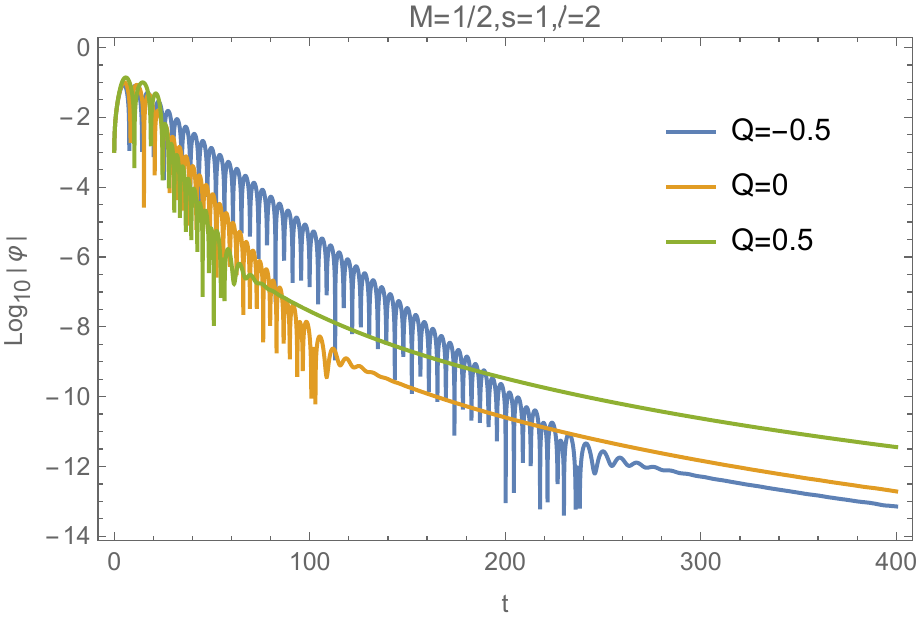}
\caption{\label{Timedomain of all matter fields1}Time evolution  for the perturbing scalar field with spin $s=0$ (left), Dirac field with $s=1/2$ (middle) and EM field with $s=1$ (right), respectively.  Here we focus on the second lowest-lying angular momentum.}
\end{figure*}

Now we move on to study the effect of the angular quantum number for various perturbations. To this end, we focus on $Q=-0.5$. The results are shown in TABLE \ref{Table:QNM grids varying l} and FIG. \ref{matterfieldsQNFl1}. Following are our observations:  (i) For small $\ell$, it has a relatively strong influence on the $Re(\omega)$, while the influence on the $Im(\omega)$ is weak. We observe that  as $\ell$ increases, $Im(\omega)$ for the EM field tends to slightly decrease, while the $Im(\omega)$ for scalar and Dirac fields slightly increase. This means that the effect of growing $\ell$ will shorten the lifetime of the EM perturbation, but it can instead extend the lifetime of the scalar and Dirac perturbations.  Similar phenomena can also be observed in the Schwarzschild black hole \cite{Zhang:2006hh},  but here we find that the introducing of the Horndeski $Q$ has print on the effect of $\ell$. In detail, positive $Q$ will enhance the effect  while negative $Q$ suppresses this effect. These findings in QNFs can be verified from FIG.\ref{Timedomain of all matter fields1} and its comparison to FIG.\ref{Timedomain of all matter fields0}.
(ii) As $\ell$ increases,  the results from WKB methods and matrix  match better and better, and the relative error becomes smaller than $10^{-6}$. This is  because the WKB-Pad\text{\'e} method is essentially a semi-analytic approximation which works better for an analytical form with higher $\ell$ \cite{Schutz:1985km}. (iii) When we further increase $\ell$, the gap among the QNFs  for all the  massless perturbing fields tends to be smaller. This is because for large $\ell$, the dominant terms in  all the  effective potentials \eqref{eq:Vsc}, \eqref{eq:Vem} and  \eqref{eq:Vdirac} are the terms $\propto\ell^2$, which have the same formula in all cases.
And finally the gap will vanish in  the eikonal limit $\ell\gg 1$ which we will study in the next subsection.
\begin{table*}[h!]
	\center
	\begin{tabular}{|c|c|c|c|c|} \hline
\null & \multicolumn{2}{c|}{ scalar field ($s=0$)} &\multicolumn{2}{c|}{relative error/\%}  \\ \hline
		$\ell$ & Matrix Method  & WKB-Pad\text{\'e} & Re($\omega$)  & Im($\omega$)  \\ \hline
  0& 0.158948 - 0.113060 i & 0.158921 - 0.113149 i & 0.0170 & -0.0787\\ \hline
  1& 0.450680 - 0.109494 i & 0.450691 - 0.109490 i & -0.0024 & 0.0037\\ \hline
  2& 0.748012 - 0.109136 i & 0.748013 - 0.109136 i & -0.0001 & $O$\\ \hline
  3& 1.046033 - 0.109036 i & 1.046033 - 0.109036 i & $O$ & $O$\\ \hline
  4& 1.344278 - 0.108994 i & 1.344278 - 0.108994 i & $O$ & $O$\\ \hline
  5& 1.642622 - 0.108973 i & 1.642622 - 0.108973 i & $O$ & $O$\\ \hline\hline
\null & \multicolumn{2}{c|}{ Dirac field ($s=1/2$)} &\multicolumn{2}{c|}{relative error/\%}  \\ \hline
		$\ell$ & Matrix Method  & WKB-Pad\text{\'e} & Re($\omega$)  & Im($\omega$)  \\ \hline
  0& 0.291568 - 0.109287 i & 0.291570 - 0.109214 i & -0.0007 & 0.0668\\ \hline
  1& 0.593509 - 0.109003 i & 0.593512 - 0.109025 i & -0.0005 & -0.0202\\ \hline
  2& 0.893211 - 0.109010 i & 0.893202 - 0.108974 i & 0.0010 & 0.0330\\ \hline
  3& 1.192322 - 0.108940 i & 1.192307 - 0.108955 i & 0.0013 & -0.0138\\ \hline
  4& 1.491178 - 0.108941 i & 1.491177 - 0.108947 i & 0.0001 & -0.0055\\ \hline
  5& 1.789936 - 0.108926 i & 1.789929 - 0.108942 i & 0.0004 & -0.0147\\ \hline\hline
\null & \multicolumn{2}{c|}{  electromagnetic field ($s=1$)} &\multicolumn{2}{c|}{relative error/\%}  \\ \hline
		$\ell$ & Matrix Method  & WKB-Pad\text{\'e} & Re($\omega$)  & Im($\omega$)  \\ \hline
  1& 0.403289 - 0.106855 i & 0.403290 - 0.106853 i & -0.0002 & 0.0019\\ \hline
  2& 0.720260 - 0.108223 i & 0.720260 - 0.108223 i & $O$ & $O$\\ \hline
  3& 1.026335 - 0.108575 i & 1.026335 - 0.108574 i & $O$ & 0.0009\\ \hline
  4& 1.328996 - 0.108716 i & 1.328996 - 0.108716 i & $O$ & $O$\\ \hline
  5& 1.630135 - 0.108787 i & 1.630135 - 0.108788 i & $O$ & -0.0009\\ \hline
  6& 1.930461 - 0.108828 i & 1.930461 - 0.108828 i & $O$ & $O$\\ \hline
	\end{tabular}
\caption{The fundamental ($n=0$) QNMs of various  massless perturbation  modes with different angular momentums obtained by WKB method and matrix method and their relative errors. Here we fix $Q=-0.5$.
\label{Table:QNM grids varying l}}
 \end{table*}

\newpage
\subsection{QNFs in eikonal limit}
 Cardoso et al proposed that in the eikonal limit ($\ell\gg 1$), the real part of QNM frequency for a static spherical black hole is connected with the angular velocity of the circular null geodesics while the imaginary part is connected with the Lyapunov exponent \cite{Cardoso:2008bp}. Then the real part of the QNMs in the eikonal limit was further related to the shadow radius of a static black hole as $\omega_{Re}=\lim\limits_{\ell\gg 1}\frac{\ell}{R_{sh}}$ \cite{Jusufi:2019ltj}, and more recently this  connection was extended into the rotating black holes \cite{Jusufi:2020dhz}.  This  correspondence may originate from the fact that the perturbing waves could be treated as massless particles propagating along the last timelike unstable orbit out to infinity, but deeper research deserves to be done for further understanding.

\begin{figure*}[htbp]
\centering
\includegraphics[height=4cm]{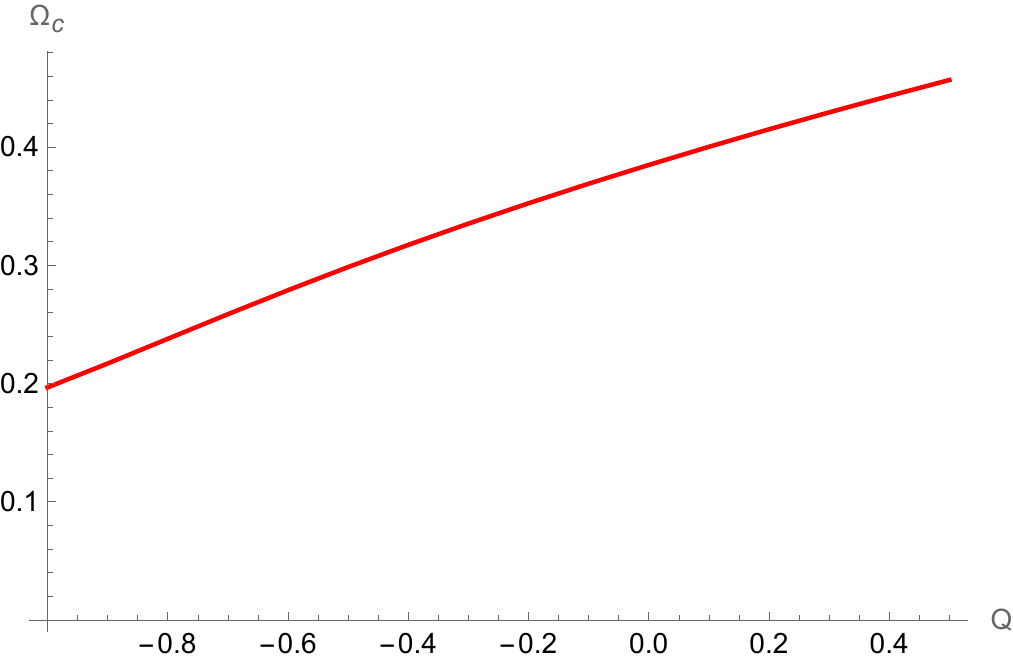}\hspace{0.5cm}
\includegraphics[height=4cm]{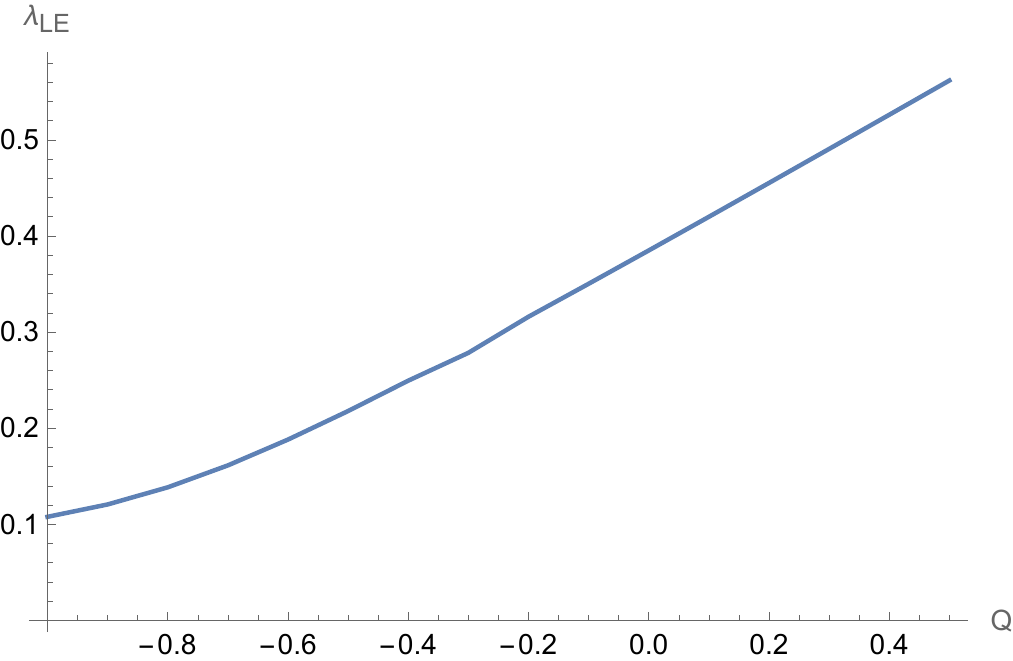}
\caption{\label{eikonalQNMfig}The angular velocity and the Lyapunov exponent as functions of hairy charge $Q$ with fixing $\ell=40$.}
\end{figure*}

In this subsection, we compare the quasinormal spectrum obtained from the master equations with the spectrum computed directly from the geometric-optics approximation formula, which is given by \cite{Cardoso:2008bp}
\begin{eqnarray}
    \omega_{QNM}&=&\Omega_{c}\ell-i(n+\frac{1}{2})|\lambda_{LE}|, \\
  \text{with} ~~~ \Omega_{c}&=&\sqrt{\frac{f(r_c)}{r^2_c}}, \text{and}~~~\lambda_{LE}=\sqrt{-\frac{r^2_c}{f(r_c)}\left(\frac{d^2}{dr^2_*}\frac{f(r)}{r^2}\right)_{r=r_c}}
\end{eqnarray}
for the Horndeski hairy black hole \eqref{eq:metric}.
 $\Omega_{c}$ is the angular velocity of a massless particle geodesically moving on a circular null orbit with radius $r=r_c$, and $\lambda_{LE}$ is the Lyapunov exponent, where the radius $r_c$ is given by the positive root of the equation $2f_c=r_cf'_c$.
In TABLE \ref{Table:eikonalQNMtable}, we list the QNFs obtained from wave analysis and the geometric-optics approximation for fixed $\ell=40$. QNFs for various perturbing fields converge to be the value obtained from geometric-optics correspondence, because in the limit $\ell\gg 1$, all the perturbed wave equations reduce to the analytical equation describing the geodesic motion of the massless particle in the Horndeski hairy spacetime.  Then in order to check the effect of $Q$, we
plot $\Omega_c$ and $\lambda_{LE}$ as functions of $Q$ in FIG. \ref{eikonalQNMfig}.  We see that both of them grow monotonically as $Q$ increases, indicating a smaller imaginary part but a larger real  part of QNFs for $\ell\gg 1$. The effect of $Q$ on the QNFs  is  already reflected for small $\ell\sim 1$ as we disclosed in the previous subsections.

\begin{table*}[h!]
	\center
	\begin{tabular}{|c|c|c|c|c|} \hline
\multicolumn{2}{|c|}{\null} &  Matrix Method  &  WKB method & geometric-optics approximation
\\\hline
\multirow{2}{*}{scalar} & Q=-0.5 & 12.090059 - 0.108933 i & 12.090062 - 0.108931 i & 11.940699 - 0.108931 i  \\ \cline{2-5}
     ~& Q=0 & 15.588767 - 0.192457 i & 15.588765 - 0.192454 i & 15.396007 - 0.192450 i  \\\hline
\multirow{2}*{EM} & Q=-0.5 & 12.088368 - 0.108929 i & 12.088371 - 0.108928 i & 11.940699 - 0.108931 i  \\\cline{2-5}
     ~ & Q=0    & 15.585599 - 0.192444 i & 15.585597 - 0.192441 i & 15.396007 - 0.192450 i  \\\hline
\multirow{2}*{Dirac} & Q=-0.5 & 12.239089 - 0.108914 i & 12.239044 - 0.108931 i   & 11.940699 - 0.108931 i  \\\cline{2-5}
      ~ & Q=0    & 15.780429 - 0.192451 i & 15.587973 - 0.192451 i   & 15.396007 - 0.192450 i  \\\hline
	\end{tabular}
 \caption{QNFs with  $\ell=40$ for various perturbing fields obtained via different methods.
 \label{Table:eikonalQNMtable}}
\end{table*}

On the other hand, the results in previous subsection show that the increasing $Q$ can enhance the contribution of the spin so that QNFs of perturbing fields with different  spins emerge a larger bifurcation (see FIG.\ref{matterfieldsQNFl0} ). However,   in the eikonal limit the perturbing fields with different spins tend to possess a same QNFs, which indicates that the contribution of field spin is diluting as $\ell$ increases. In order to analytically understand the balance effect of $Q$ and $\ell$ on
the spin contribution on the QNFs, we apply the Newman-Penrose formalism to construct general spherical symmetric Teukolsky equations for an arbitrary field spin $s$ in static spherical-symmetric metric with $ds^2=-f(r)dt^2+f^{-1}(r) dr^2+r^2(d\theta^2+\sin^2\theta d\varphi^2)$, which separate into  angular and radial  parts as
\begin{eqnarray}
\label{Eq:teusphs1}
    \left[\frac{1}{sin\theta}\frac{d}{d\theta} \left( sin\theta \frac{d}{d\theta} \right) -\frac{m^2+2m\,s\,cos\theta+s^2cos^2\theta}{sin^2\theta}+s+A_{s\ell} \right]S_s(\theta)=0,\\
\label{Eq:teusphr1}
   \left[ \Delta^{-s}\frac{d}{dr}\left(\Delta^{(1+s)}\frac{d}{dr} \right) +
4i\,s\,r\,\omega +\frac{r^2\omega(r^2\omega-i\,s\,\Delta')}{\Delta}+\epsilon_s(\Delta''-2)-A_{s\ell}  \right]R_s(r)=0.
\end{eqnarray}
with $\Delta\equiv r^2f(r)$, $m$  the azimuthal number,  $\epsilon_s$ and $A_{s\ell}$ listed in TABLE \ref{table:l-ac}. For more details on the formula derivation, readers can refer to \cite{Teukolsky:1973ha, Jing:2005ux, Harris:2003eg}.  It is straightforward to check that the above equations can reduce to the Teukolsky equation in static limit derived in \cite{Teukolsky:1973ha}.
\begin{table}[H]
\center{
\begin{tabular}{|c| c |c| c| c| c |c |c |c|}
  \hline
 $s$ & $-2$ & $-1$ & $-1/2$ & $0$ & $1/2$ & $1$ & $2$  \\ \hline
 $\epsilon_s$  & $1/2$ & $0$ &$0$ &$0$ & $1/2$& $1$ & $5/2$  \\ \hline
 $A_{s\ell}$  & $\quad (\ell-1)(\ell+2)$ & $\quad \ell(\ell+1)\quad $ &$\quad \ell^2 \quad$ &$\qquad \ell(\ell+1) \qquad$ & $\; (\ell+1)^2-1 \;$& $\quad \ell(\ell+1)-2 \quad$ & $\quad (\ell-1)(\ell+2)-4 \quad$  \\
  \hline
\end{tabular}
\caption{Variables $A_{S\ell}\,\& \, \epsilon_s$ dependent on the field spin in spherical symmetric Teukolsky equations.}\label{table:l-ac}}
\end{table}

To proceed, we reform $R_s=\Delta^{-s/2} \Psi_s/r$ and work in  tortoise coordinate  $dr_*=(r^2/\Delta)dr$ , thus  the radial equation \eqref{Eq:teusphr1} is transformed into the wave-like equation
\begin{eqnarray}
\label{Eq.teusphr2}
 \frac{d^2\Psi_s}{dr_*^2}&+&[\omega^2-V_{s\ell}(r)]\Psi_s=0,  ~~~
 \text{with}\\
 \label{Eq.teupotential}
 V_{s\ell}(r)=i\,s\,\omega \, r^2\frac{d}{dr}\left( \frac{\Delta}{r^4}\right) &+& \frac{\Delta}{r^3}\frac{d}{dr}\left( \frac{\Delta}{r^2}\right)+\frac{\Delta}{4r^4} \left(4A_{s\ell}+s^2\frac{\Delta'^2}{\Delta}+2s\Delta''- 4\epsilon_s(\Delta''-2)\right).
\end{eqnarray}
It is noted that the Teukolsky radial equations are different from the corresponding master equations \eqref{eq:scalar master eq}, \eqref{eq:EM master eq} and \eqref{eq:Dirac master eq} when setting $s=0,1,1/2$, because they are not in the form of canonical wave equations. However, it was addressed in \cite{Chandrasekhar:1975nkd} that the Teukolsky equations can be brought into the corresponding master equations under certain transformation.
The QNFs are significantly determined by the effective potential and dominated by the last term in $V_{s\ell}$. In the Horndeski hairy black hole \eqref{eq:metric},  the influence from the spin can be amplified through the coupling $s^2Q^2$ originating from the term $s^2\Delta'^2/4r^4$, which eventually causes the bifurcation or, saying the wider ``fine structure" in the quasinormal spectrum. While in the eikonal limit  $\ell>>1$, the term  $\Delta A_{s\ell}/r^4 \rightarrow \Delta \ell^2/r^2$ shall become dominant in the potential, so that  the effect of spin will be suppressed. This subsequently causes the degeneracy for both the null circular orbits and the quasinormal spectrum of massless fields with different spins.

\section{Greybody factor and Hawking radiation}\label{sec:Hawking Radiation}

It is known from Hawking's paper  \cite{Hawking:1975vcx} that a particle with negative energy measured by an observer at infinity can physically exist inside the black hole since there the Killing vector  is spacelike. Thus, during the pair production at the vicinity of the horizon,  the particle with positive energy can escape to the observer and leave the negative one to fall into the singularity. This effect causes the so-called Hawking radiation, whose power spectrum is shown to be a literally black-body spectrum.  However, due to the existence of the potential barrier outside the black hole, the Hawking radiation is not totally transparent for the observer at infinity, so what the observer detects in fact is a grey-body spectrum because the particles could be scattered by the potential barrier.

In order to quantize the scattering process for various particles, one should first calculate the transmission coefficient, which is also defined as the greybody factor. Here we will  solve the master equations \eqref{eq:scalar master eq},  \eqref{eq:EM master eq} and \eqref{eq:Dirac master eq} by the scattering boundary conditions which permit the ingoing wave at infinity differing from the case of calculating QNFs. The boundary conditions that can equivalently describe the scattering process for a particle emitted from the horizon read as
\begin{equation}
\begin{split}
    \Psi_{\ell}&=T_{\ell}\, e^{-i\omega r_*},\qquad \quad r_*\rightarrow -\infty,\\
     \Psi_{\ell}&=e^{-i\omega r_*}+R_{\ell}\, e^{i\omega r_*}, \, \qquad \quad r_*\rightarrow +\infty,\\
\end{split}
\end{equation}
where $T_{\ell}$ and $R_{\ell}$ are denoted as the transmission and reflection coefficients for the angular momentum $\ell$ mode, satisfying  $|T_{\ell}|^2+|R_{\ell}|^2=1$.  Since from section \ref{sec:eoms} we know that each effective potential exhibits a potential barrier  decreasing monotonically towards both boundaries, thereby, we can again employ  the WKB method described in appendix \ref{sec:appendix1} to calculate the coefficients.
Subsequently, we have the greyfactor $|A_{\ell}|^2$ defined as \cite{Schutz:1985km,PhysRevD.35.3621}
\begin{equation}
 |A_{\ell}|^2=1-|R_{\ell}|^2=|T_{\ell}|^2~~~\text{and}~~~    R_{\ell}=(1+e^{-2i\pi \mathcal{K}})^{-\frac{1}{2}}
\end{equation}
where $\mathcal{K}$ can be obtained from the WKB formula
\begin{eqnarray}
    \mathcal{K}-i \frac{\omega^2-V(r_0)}{\sqrt{-2V''(r_0)}}-\sum_{i=2}^{i=6} \Lambda_i(\mathcal{K}) =0.
\end{eqnarray}
Here, $V(r_0)$ denotes the maximal potential locating at  $r_0$  for various spins, the second derivative in $V$ is with respect to the tortoise coordinate, and $\Lambda_i$ are  the higher order WKB correction terms.  It is noted that the WKB formula for determining the  grey-body factors is well known to provide reasonable accuracy for further estimating the energy rate of Hawking radiation. {However, this approach may not be suitable for the cases with very small $\omega$, which
imply  almost complete wave reflection with negligible contributions to the total
energy emission rate, so we set a cutoff for $\omega$ in the numeric to make our results reliable.}
\begin{figure*}[htbp]
\centering
\includegraphics[height=3.2cm]{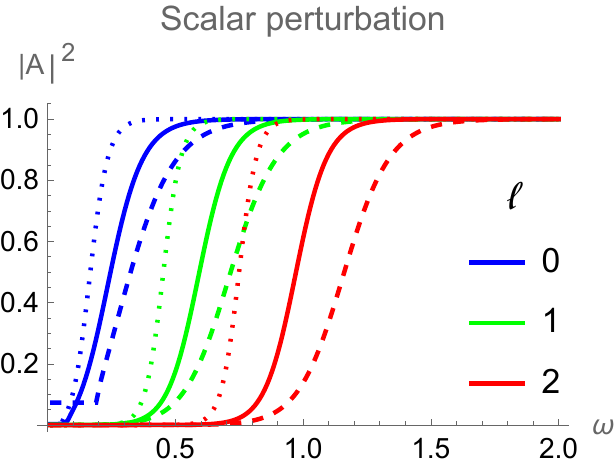}\hspace{0.5cm}
\includegraphics[height=3.2cm]{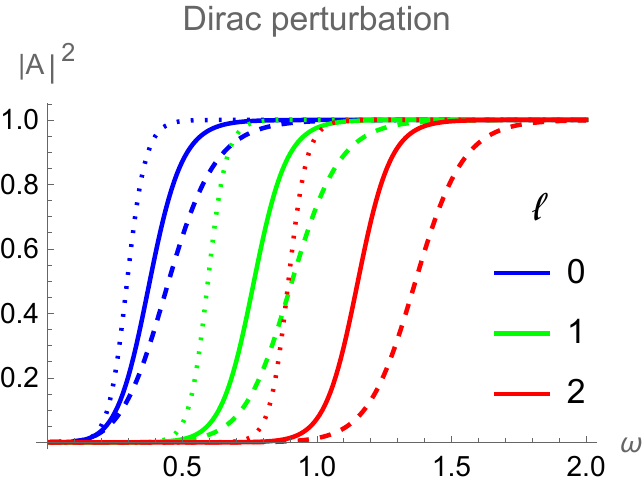}\hspace{0.5cm}
\includegraphics[height=3.2cm]{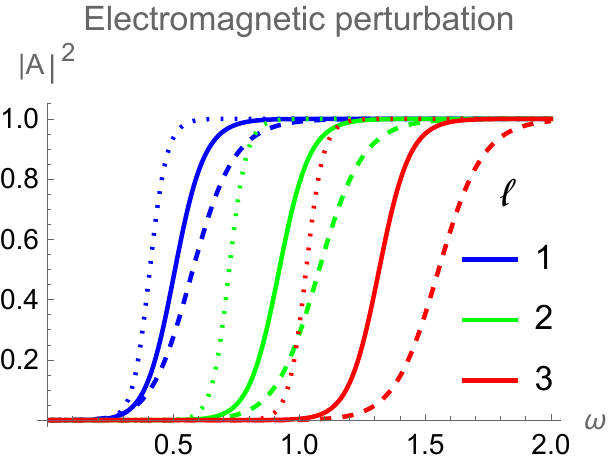}
\caption{\label{graybodyA}Greybody factors for emission particles with spin $0$ (left), $1/2$ (middle) and $1$ (right). In each plot, we distinguish the curves with different $\ell$ by colors, while curves with same color indicate the graybody factor with $Q=-0.5 $ (dotted), $Q=0$ (solid),$Q=0.5$ (dashed).}
\end{figure*}

With the above preparation of methodology, we depict the grey-body factor as a function of $\omega$ with different Horndeski hair $Q$ for various fields in FIG. \ref{graybodyA}.  Two remarkable properties we can extract from the figure. (i) Similar as in Schwarzschild black hole ($Q=0$),  the increasing  of spin $s$  and orbital angular momentum $\ell$ could make the curves of the grey-body factor shift towards a larger $\omega$ in Horndeski hairy black hole. This means that for the emission particle with larger $\ell$ and $s$, the lowest {frequency} with which the particle  penetrates the potential barrier would raise up such that the barrier would tend to shield more low-frequency particles for the observer at infinity.
(ii) Regarding to the effect of the hairy charge $Q$ on $|A_\ell|^2$ ,  we observe that the increasing of $|Q|$ would not make the curve of $|A_\ell|^2$ shift distinctly, instead, it makes the curve branch out between two fixed frequencies. It means that the Horndeski hair would not change the lowest {frequency} for a particle transmitting the potential barrier. Moreover, by comparing the dotted, solid and dashed curves with the same color for all the fields, it is obvious that for larger $Q$, the grey-body factors is smaller for fixed $\omega$, which means that more number of particles is reflected by the corresponding effective potential. This observation is reasonable because for larger $Q$, all the effective potentials barrier is higher (see FIG.\ref{Fig:Vsc}-FIG.\ref{fig:V_Dirac}), so the particles  are more difficult to penetrate.

With the greybody factor in hands, we can further estimate the energy emission rate of the Hawking radiation of the Horndeski hariy black hole via \cite{Hawking:1975vcx}
\begin{eqnarray}
\label{eq:EER}
    \frac{dE}{dt}=\sum_{\ell}\frac{N_{\ell}}{2\pi}\frac{|A_{\ell}|^2}{e^{\omega /T_H}\pm 1}\omega d\omega,
\end{eqnarray}
 where $\pm$  in the denominator denote the fermions and the bosons created in the vicinity of the event horizon, the Hawking temperature $T_H$ is
 \begin{eqnarray}
 \label{Eq:HawkingTemperature}
     T_H=\frac{f'(r)}{4\pi}\big |_{r=r_+=2M}=\frac{1+Q}{8\pi M},
 \end{eqnarray}
 and the multiples $N_{\ell}$  is known as
\begin{eqnarray}
N_{\ell}=\left\{ \begin{aligned}
&2\ell+1  &\text{(scalar field),}\quad \\
&8(\ell+1)  &\text{(Dirac field),}\quad \\
&2(2\ell+1)    &\text{(Maxwell field).}
\end{aligned} \right.
\end{eqnarray}
Note that the formula \eqref{eq:EER} only works when the system can be described by the canonical ensemble, which is fulfilled under the assumption that the temperature of black hole does not change between two particles emitted in succession \cite{Hawking:1975vcx}.

\begin{figure*}[htbp]
\includegraphics[height=3cm]{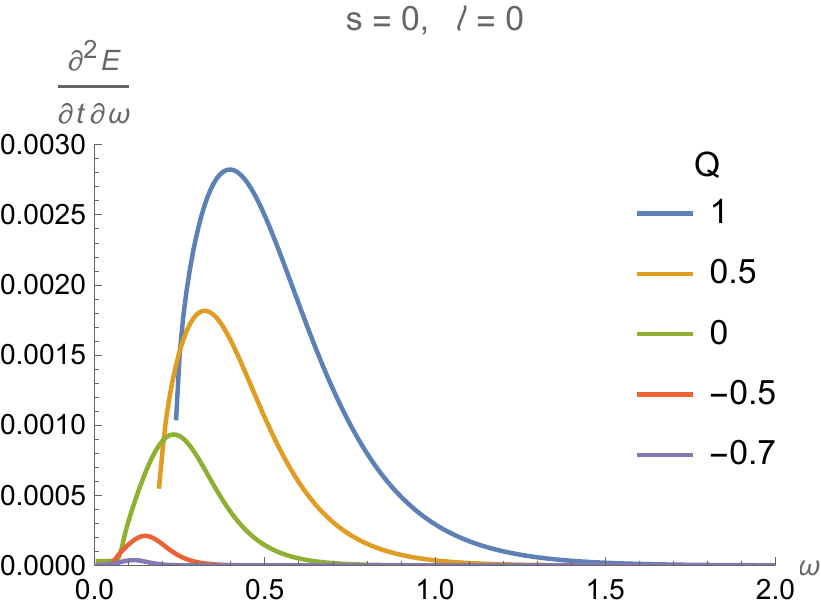}\hspace{0.5cm}
\includegraphics[height=3cm]{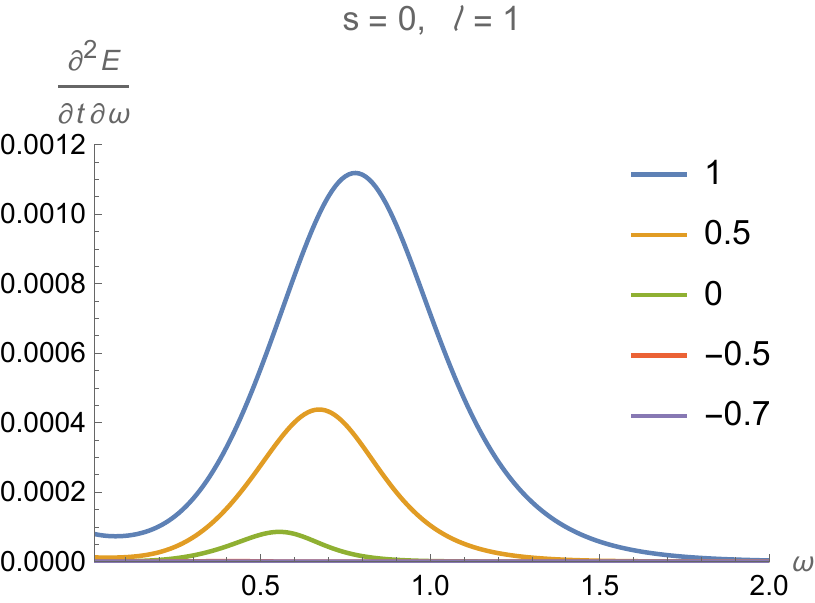}\\
\includegraphics[height=3cm]{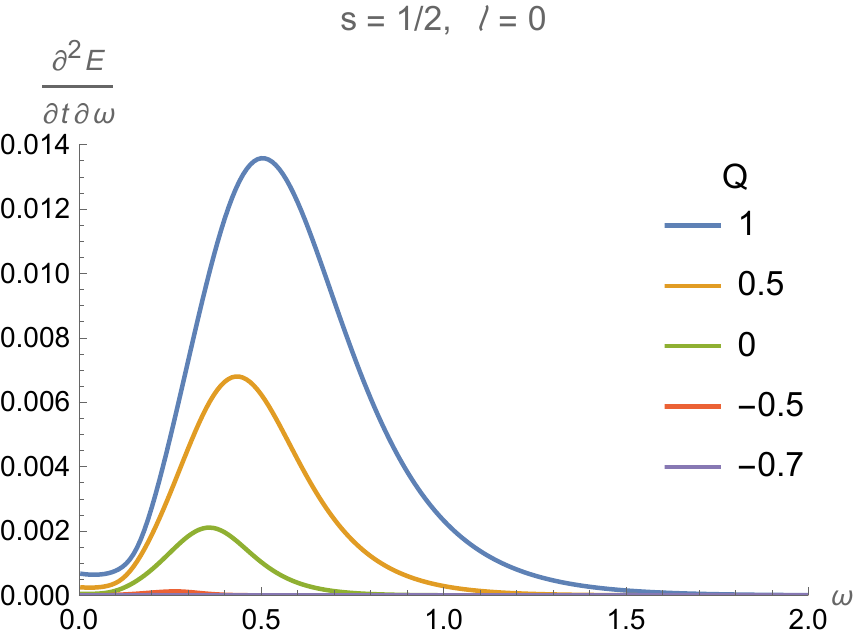}\hspace{0.5cm}
\includegraphics[height=3cm]{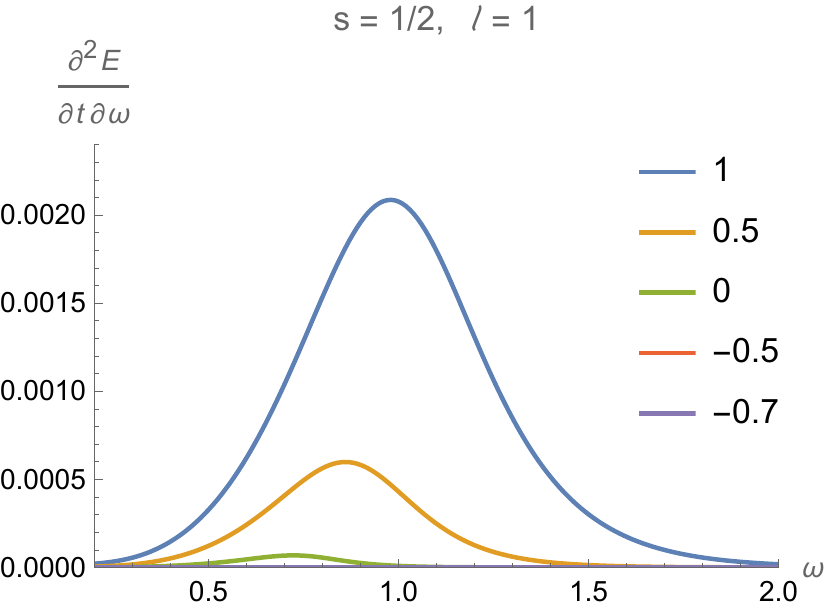}\\
\includegraphics[height=3cm]{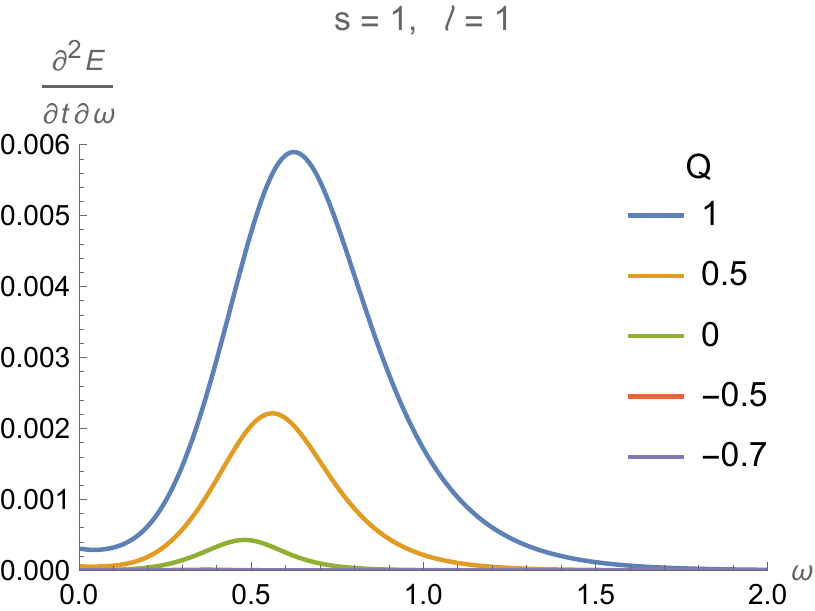}\hspace{0.5cm}
\includegraphics[height=3cm]{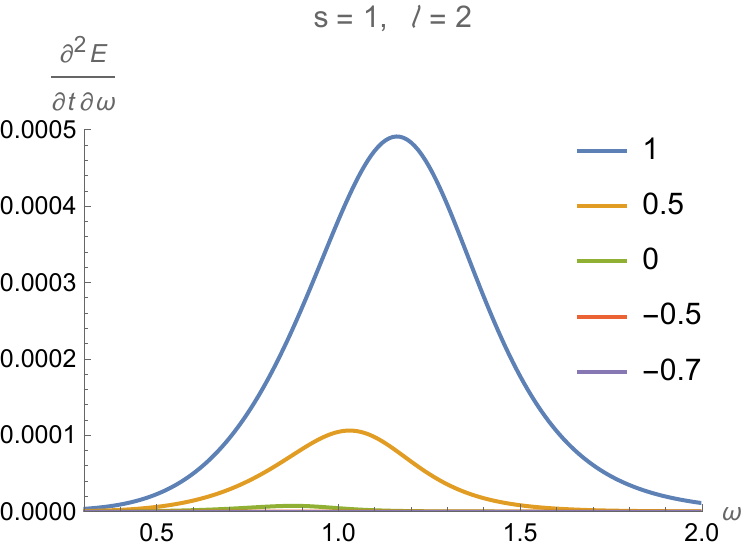}\\
\caption{\label{EERperw total}Energy emission rate as a function of frequency for various fields with spin $s=0$ (top), $s=1/2$ (middle), $s=1$ (bottom) around Horndeski hairy black hole. The left column describes the fields with the corresponding lowest $\ell$ while the right column  is for their second lowest $\ell$.}
\end{figure*}

The energy emission rate (EER) for scalar, Dirac and  Maxwell fields as a function of frequency with samples of $\ell$ and $Q$  are shown in Fig.\ref{EERperw total}. We can extract the following features.
(i) The general behavior of EER is that as $\omega$ grows, the EER  first increases till it reaches a  maximum at $\omega=\omega_p$ where $\omega_p$ depends on the parameters, and then decays to be zero. This is because in the right side of \eqref{eq:EER}, there exists a competition between the  greybody factor in the numerator and the exponential term in the denominator.  Though FIG.\ref{graybodyA} shows that $|A_\ell|^2$ grows as {$\omega$ increases},  the growth of $|A_\ell|^2$ is only more influential for $\omega<\omega_p$, while for $\omega>\omega_p$ the growing of $e^{\omega/T_H}$ plays the dominant role and makes the EER decrease. And further increasing $\omega$,  $|A_\ell|^2$ tends to the unit, but the denominator increases exponentially, which causes the EER to decay exponentially.
(ii) In each plot,  we see that the  EER for various fields in the black hole with a larger $Q$ is stronger. The result is reasonable because for fixed $\omega$, as $Q$ increases,  the greybody factor becomes smaller and the exponential term in the denominator also decreases due to the growth of Hawking temperature \eqref{Eq:HawkingTemperature}. The joint contributions result in the stronger intensity of EER for larger $Q$. (iii) By comparing the two plots for each field, we observe that EER for particles with higher orbital angular momentum would be suppressed in both Schwarzschild and Honrdeski hairy black holes. In addition, this suppression effect is more significant for the particles with a higher spin.

\begin{figure*}[htbp]
\centering
\includegraphics[height=3.2cm]{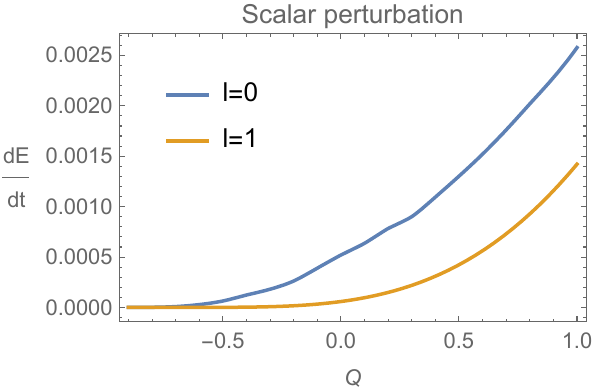}\hspace{0.5cm}
\includegraphics[height=3.2cm]{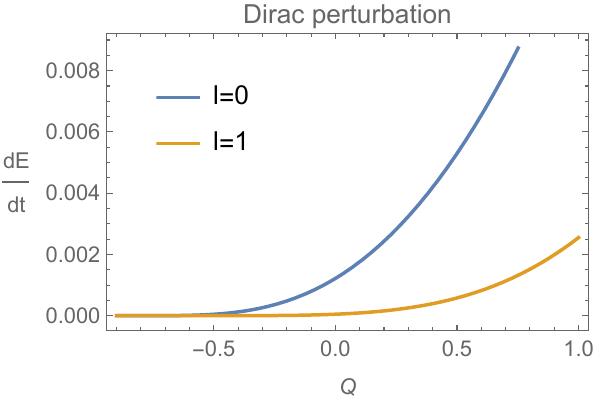}\hspace{0.5cm}
\includegraphics[height=3.2cm]{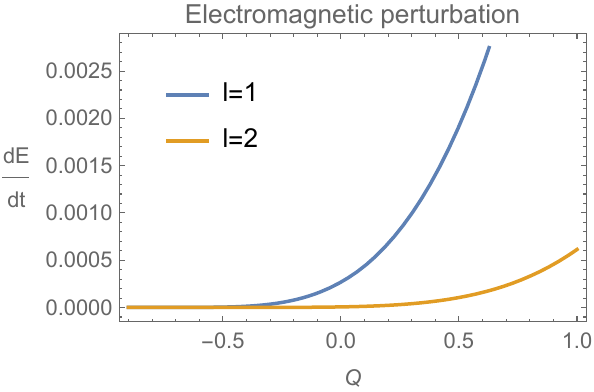}
\caption{\label{EER total}Total emission rate $dE/dt$ as a function of Horndeski hair for various fields with $s=0$ (left), $s= 1/2$ (middle) and $s= 1$ (right).}
\end{figure*}

Finally, we numerically integrate the EER over $\omega$ and obtain the total EER, $dE/dt$, for various fields as the function of $Q$ in FIG.\ref{EER total}. For the static hairy black hole in the current Horndeski gravity, the positive  Horndeski hair $Q$ will enhance the total EER around the black hole, which may lead to a higher speed of evaporation and a shorter lifetime for this kind of black hole according to the discussion in \cite{PhysRevD.13.198}.
Therefore,  in terms of observation, this intuitively would mean that, such hairy small or even medium black hole with a large positive $Q$ in the early universe may have disappeared due to the high evaporation rate. While for negative $Q$, one would expect that the Horndeski hairy black hole has a lower evaporation rate. Especially, in the extremal case with $Q=-1$, the total EER  approaches zero as expected, meaning that such extremal black holes almost do not evaporate and thus they are often considered to live forever if they are in complete isolation, similar to the case in extremal RN black hole \cite{Adams:2000ax}.

\section{Conclusion and discussion}\label{sec:conclusion}

In this paper,  we investigated the quasinormal frequencies  and Hawking radiations of the Horndeski hairy black hole by analyzing the perturbations of massless fields with spins $0$ (scalar field), $1/2$ (Dirac field) and $1$ (electromagnetic field), respectively. The starting points of both aspects  are the master equations of the perturbing fields.  For the quasinormal mode spectral analysis, we employed three methods: WKB method, matrix method and time domain integration; while in the Hawking radiation part, we used the $6-$th order WKB method to determine the greybody factor.

Our results show that under the massless perturbations of the scalar field, Dirac field and electromagnetic field, the Horndeski hairy black holes are dynamically stable in terms of the frequency and time domain of QNMs. Similar as in Schwarzschild  black hole \cite{Kokkotas:1999bd}, the massless field with higher spin has a larger imaginary part of quasinormal frequency, so the related perturbation can live longer in the hairy  black hole background. This effect can be enhanced by having a stronger negative Horndeski hair. Moreover, the real part of QNFs for all the perturbations increases as $Q$ increases, meaning that the larger Horndeski hair  enhances the oscillation of the perturbations.
In addition, in Horndeski hairy black hole,  the mode with a larger $\ell$ has a shorter lifetime for electromagnetic field  perturbation, but it can survive longer for the scalar and Dirac perturbations, which are similar to those in Schwarzschild black hole. But this effect of $\ell$ on the QNFs would be enhanced by
a positive $Q$ while suppressed by a negative $Q$. Nevertheless, for large enough $\ell$ till the eikonal limit $(\ell\gg 1)$,  the QNFs for various perturbations tend to be almost the same value, of which the imaginary (real) part decreases (increases) as $Q$ increases.  Focusing on the  general Teukolsky equations for arbitrary spin in the current hairy black hole,  we explained  the balance effects of the Horndeski hair, the spin and the quantum angular momentum on the QNFs in an analytical way.

Our studies on Hawking radiation shew that the intensity of energy emission rate for various fields is stronger for a larger Horndeski hair,  which indicates that the Horndeski hair  could remarkably influence the evaporation rate of  black hole. Thus, we argue that such  black hole with large positive Horndeski hairy charge (if exists) in the early universe may have disappeared due to the strong radiation, while comparing to black hole in GR, Horndeski hairy black hole with negative hair could have a lower evaporation rate.

It would be more practical to extend our study into the gravitational perturbations, i.e., the massless spin-2 field, and we believe that it deserves an individual consideration due to the  difficulty in reducing the master wavelike equation in the current theory. However,  our findings from the test external fields could shed some insights on the related physics on gravitational perturbations. For example,  the QNFs of gravitational fields perturbation in eikonal limit could be the same as our findings because  the behavior of the QNM spectrum for external and gravitational fields is usually known to be qualitatively the same, independent of the spin of the field in this limit. In addition, our findings about the influence of the field's spin on the QNFs and the energy emission rate could also provide a good reference in this scenario.

Moreover, considering that the overtone modes may play important roles in the GW \cite{Giesler:2019uxc, Sago:2021gbq}, it would be interesting to further study the QNFs for the overtone modes.  Additionally, the extension to the external fields with mass could be another interesting direction.  At least the propagation of the massive scalar field was found to behave differently from the massless scalar field \cite{Konoplya:2004wg, Tattersall:2018nve}. We hope to perform these studies in the near future.

\begin{acknowledgments}
We appreciate Guo-Yang Fu and Hua-Jie Gong for helpful discussions.
This work is partly supported by Natural Science Foundation of China under Grants No. 12375054 and No. 12375055, Natural Science Foundation of Jiangsu Province under Grant No.BK20211601, the Postgraduate Research \& Practice Innovation Program of Jiangsu Province under Grant No. KYCX23\_3501, and Top Talent Support Program from Yangzhou University.
\end{acknowledgments}

\begin{appendices}
\section{WKB method}\label{sec:appendix1}
WKB method, as a well known approximation, is a semianalytic technique for determining the eigenvalue of the Schrodinger wavelike equation, which has the form:
\begin{equation}
 \frac{d^2}{dx^2}\psi(x)-V(x)\psi(x)=0,
\end{equation}
where the potential barrier $V(x)$ rises to a peak at $x=x_0$ and is assumed to be constant at the infinity $(|x| \to \infty)$, and the radial function $\psi(x)$ is required to be purely ``outgoing" as $|x| \to \infty$.
The method is named after Wentzel-Kramers-Brillouin, and originally applied to approximate the bound-state energies and tunneling rates of the Schrodinger equation in quantum physics. Owing to the noted modification of the WKB approach proposed by Iyer and Will \cite{PhysRevD.35.3621}, the method is carried to the third order beyond the eikonal approximation and is able to calculate the quasi-normal frequency quickly and accurately for a wide range of black hole systems. Then Konoplya extended the method to the 6th order \cite{Konoplya:2003ii}, and Matyjasek-Opala brought it to the 13th order\cite{Matyjasek:2017psv}. The principal idea is to match simultaneously exterior WKB solutions across the two turning points on the potential barrier, and this finally yields the WKB formula
\begin{equation}
\label{WKBmastereq}
    \frac{i V(x_0)}{\sqrt{2V''(x_0)}}-\sum_{i=2}^{N} \Lambda_i =n+\frac{1}{2},
\end{equation}
where $n=0,1,2,...$ is the overtone number and $N$ is the number of WKB order. $\Lambda_i$ is the $i$-th correction term that depends only on the derivatives of $V(x)$ evaluated at $x_0$, and several  formulas can be found in \cite{PhysRevD.35.3621,Konoplya:2003ii}. In our framework, we substitute the $V(x)$ by  $V_{sc}(r),  V_{EM}(r)$ and $V_{Dirac}(r)$ for the scalar, electromagnetic and Dirac fields, respectively, into \eqref{WKBmastereq}, and then solve the equations to obtain the QNFs.

\section{Matrix method}\label{sec:appendix2}
In this appendix, we will show the main steps of matrix method \cite{Lin:2016sch,Lin:2017oag,Lin:2019mmf} in calculating the QNFs from the three master equations for various perturbing fields.  We uniform the three  master equations as
\begin{equation}\label{eq:radial eq}
\frac{d^2K}{dr_*^2}+[\omega^2-V(r)]K=0,
\end{equation}
where $K$ is the corresponding field variables, and the effective potential $V(r)$ can be $V_{sc}(r),V_{EM}(r)$ or $V_{Dirac}(r)$ for the scalar, electromagnetic and Dirac field, respectively. To study the QNM spectrum, we impose the ingoing wave  $K\sim e^{-i\omega r_*}$ at horizon $(r\to r_+)$, and the outgoing wave $K\sim e^{i\omega r_*}$ at infinity  $(r\to\infty) $.
Recalling \eqref{eq:rstar}, the above boundary conditions can be rewritten as
\begin{equation}
K(r\to r_+)\sim
\begin{cases}
(r-r_+)^{-\frac{i\omega r_+^2}{Q+r_+}},\quad\text{for non--extreme black hole ($Q>-2M$)}\\
\\
(r-r_+)^{-\frac{10}{3}i\omega r_+}e^{\frac{2i\omega r_+^2}{r-r_+}},\quad\text{for extreme black hole ($Q=-2M$)}
\end{cases},
\end{equation}
and
\begin{equation}
K(r\to\infty)\sim r^{2iM\omega}e^{i\omega r-\frac{1}{2}iQ\omega(\ln\frac{r}{2M})^2}.
\end{equation}
Thus, in order to make sure $K$ satisfy the boundary conditions simultaneously, it is natural to redefine $K$ in the following way
\begin{equation}\label{eq:KtoZ}
K(r)=
\begin{cases}
(r-r_+)^{-\frac{i\omega r_+^2}{Q+r_+}}r^{\frac{i\omega r_+^2}{Q+r_+}}r^{2iM\omega}e^{i\omega r-\frac{1}{2}iQ\omega(\ln\frac{r}{2M})^2}Z(r),\quad\text{for non--extreme black hole}\\
\\
(r-r_+)^{-\frac{10}{3}i\omega r_+}e^{\frac{2i\omega r_+^2}{r-r_+}}r^{\frac{10}{3}i\omega r_+}e^{-\frac{2i\omega r_+^2}{r}}r^{2iM\omega}e^{i\omega r-\frac{1}{2}iQ\omega(\ln\frac{r}{2M})^2}Z(r),\quad\text{for extreme black hole}
\end{cases}.
\end{equation}

To proceed, we consider the coordinate transformation
\begin{eqnarray}\label{eq:rtox}
      x(r)=
        \begin{cases}
           1-(r_+/r)^{1/3},\quad\text{for scalar and electromagnetic field}\\
            \\
         (1-(r_+/r)^{1/2})^{1/2},\quad\text{for Dirac field}
        \end{cases} ,  ~~~~~
    \label{obser2}
\end{eqnarray}
to bring the integration domain from $r\in[r_+,\infty]$ to $x\in(0,1]$. Further by implementing the function transformation
\begin{equation}\label{eq:Ztochi}
\chi(x)=x(1-x)Z(x),
\end{equation}
we can reform the equation \eqref{eq:radial eq} into
\begin{equation}\label{eq:chi}
A_2(x)\chi''(x)+A_1(x,\omega)\chi'(x)+A_0(x,\omega)\chi(x)=0,
\end{equation}
where the expressions of $A_i(i=0,1,2)$ are straightforward and will  not present here. Subsequently, the boundary conditions are  simplified as
\begin{equation}\label{eq:bc chi}
\chi(0)=\chi(1)=0.
\end{equation}

With the reformed master equation \eqref{eq:chi} and the boundary conditions \eqref{eq:bc chi} in hands, we can follow the standard steps of matrix method  directly to obtain the eigenvalue $\omega$ in it. Following is the main principles of  the matrix method  \cite{Lin:2016sch,Lin:2017oag,Lin:2019mmf}.  One first interpolates $N$ grid points $x_1=0<x_2<x_3<\dots<x_N=1$ in the interval $x\in[0,1]$, then by carrying out Taylor expansion of $\chi$ at anyone of these grid points $x_k(k=1,2,\cdots,N)$, one has
\begin{equation}
\chi(x)-\chi(x_k)=(x-x_k)\chi'(x_k)+\frac{1}{2}(x-x_k)^2\chi''(x_k)+\frac{1}{3!}(x-x_k)^3\chi'''(x_k)+\cdots,
\end{equation}
Finally, by taking $x=x_j(j=1,2,\cdots,k-1,k+1,\cdots,N)$, one can get a matrix equation
\begin{equation}
\Delta F=MD,
\end{equation}
where
\begin{equation*}
\Delta F=\left(\chi(x_1)-\chi(x_k),\chi(x_2)-\chi(x_k),\cdots,\chi(x_{k-1})-\chi(x_k),\chi(x_{k+1})-\chi(x_k),\cdots,\chi(x_N)-\chi(x_k)\right)^T,
\end{equation*}
\begin{equation*}
M=\begin{pmatrix}
x_1-x_k&\frac{(x_1-x_k)^2}{2}&\cdots&\frac{(x_1-x_k)^k}{k!}&\cdots&\frac{(x_1-x_k)^{N-1}}{(N-1)!}\\
x_2-x_k&\frac{(x_2-x_k)^2}{2}&\cdots&\frac{(x_2-x_k)^k}{k!}&\cdots&\frac{(x_2-x_k)^{N-1}}{(N-1)!}\\
\cdots&\cdots&\cdots&\cdots&\cdots&\cdots\\
x_{k-1}-x_k&\frac{(x_{k-1}-x_k)^2}{2}&\cdots&\frac{(x_{k-1}-x_k)^k}{k!}&\cdots&\frac{(x_{k-1}-x_k)^{N-1}}{(N-1)!}\\
x_{k+1}-x_k&\frac{(x_{k+1}-x_k)^2}{2}&\cdots&\frac{(x_{k+1}-x_k)^k}{k!}&\cdots&\frac{(x_{k+1}-x_k)^{N-1}}{(N-1)!}\\
\cdots&\cdots&\cdots&\cdots&\cdots&\cdots\\
x_N-x_k&\frac{(x_N-x_k)^2}{2}&\cdots&\frac{(x_N-x_k)^k}{k!}&\cdots&\frac{(x_N-x_k)^{N-1}}{(N-1)!}
\end{pmatrix},
\end{equation*}
\begin{equation*}
D=\left(\chi'(x_k),\chi''(x_k),\cdots,\chi^{(k)}(x_k),\cdots,\chi^{(N-1)}(x_k)\right)^T.
\end{equation*}
It is more convenient to express $\chi'(x_k)$ and $\chi''(x_k)$ as the linear combination of the function value of $\chi$ at each grid point using  the Cramer rule,
\begin{equation}
\begin{split}
\chi'(x_k)&=\det(M_1)/\det(M),\\
\chi''(x_k)&=\det(M_2)/\det(M),
\end{split}
\end{equation}
where the matrix $M_i (i=1,2)$ is constructed with replacing the $i$'th column of matrix $M$ by $\Delta F$. In this way,  one can finally transform the master equation \eqref{eq:chi} into a matrix equation
\begin{equation}\label{eq:Mbar}
\bar{\mathcal{M}}(\omega)\mathcal{F}=0,
\end{equation}
where $\mathcal{F}=(\chi(x_1),\chi(x_2),\dots,\chi(x_N))^T$.  Considering  the boundary conditions \eqref{eq:bc chi} , the above matrix equation take the forms
\begin{equation}\label{eq:M}
\mathcal{M}(\omega)\mathcal{F}=0,
\end{equation}
where
\begin{equation}
\mathcal{M}_{ij}=\bigg\{
\begin{aligned}
\delta_{ij},\quad i&=1,N\\
\bar{\mathcal{M}}_{ij},\quad i&=2,3,\dots,N-1
\end{aligned}.
\end{equation}
Consequently, the condition that Eq.\eqref{eq:M} has nonvanishing root is the validity of the algebra equation
\begin{equation}
det(\mathcal{M}(\omega))=0,
\end{equation}
by solving which, one obtains the eigenvalue $\omega$ as the quasinormal frequencies.

\section{Time domain integration}\label{sec:appendix3}
In order to illustrate the properties of QNMs from the propagations of various fields, we shall  shift our analysis into  the time domain. To this end, we reconstruct the Schrodinger-like equations (Eqs. \eqref{eq:scalar master eq}, \eqref{eq:EM master eq}  and \eqref{eq:Dirac master eq}) into the time-dependent form by simply replacing the $\omega^2$ with $-d^2/dt^2$, then we have the uniformed second-order partial differential equation for various perturbed fields as
\begin{eqnarray}\label{eq:Time-eq}
    \left(-\frac{d^2}{dt^2}+\frac{d^2}{dr_*}-V(r) \right)\Psi(t,r)=0.
\end{eqnarray}

To solve the above equations,  one has to deal with the time-dependent evolution problem. A convenient way is to adopt the finite difference method \cite{Abdalla:2010nq} to numerically integrate these wave-like equations at the time coordinate and fix the space configuration with a Gaussian wave as an initial value of time. To handle this, one firstly discretizes the radial coordinate with the use of the definition of tortoise coordinate
\begin{eqnarray}
    \frac{dr(r_*)}{dr_*}=f(r(r*)) \Rightarrow \frac{r(r_{*j}+\Delta r_*)-r(r_{*j})}{\Delta r_*}=\frac{r_{j+1}-r_j}{\Delta r_*}=f(r_j) \Rightarrow r_{j+1}=r_j +\Delta r_* f(r_j),
\end{eqnarray}
So, a list of $\{ r_j \}$ is generated  if one  chooses the seed $r_0=r_{horizon}+\epsilon$ with a given the grid interval $\Delta r_*$. Then one can further discretize the effective potential into $V(r(r_*))=V(j\Delta r_*)\equiv V_j$ and the field into $\Psi(t,r)=\Psi(j\Delta r_*,\, i \Delta t)\equiv \Psi_{j,i}$. Subsequently,  the wave-like equation \eqref{eq:Time-eq} turns out to be a discretized equation
\begin{eqnarray}
    -\frac{\Psi_{j,i+1}-2\Psi_{j,i}+\Psi_{j,i-1}}{\Delta t^2}+\frac{\Psi_{j+1,i}-2\Psi_{j,i}+\Psi_{j-1,i}}{\Delta r_*^2}-V_j \Psi_{j,i}+\mathcal{O}(\Delta t^2)+\mathcal{O}(\Delta r_*^2)=0,
\end{eqnarray}
from which one can isolate $\Psi_{j,i+1}$ after algebraic operations
\begin{eqnarray}
    \Psi_{j,i+1}=\frac{\Delta t^2}{\Delta r_*^2} \Psi_{j+1,i}+\left(2-2\frac{\Delta t^2}{\Delta r_*^2}- \Delta t^2 V_j \right) \Psi_{j,i}+\Psi_{j-1,i}-\Psi_{j,i-1}.
\end{eqnarray}
The above equation is nothing but an iterative equation, which can be solved if one gives a Gaussian wave packet $\Psi_{j,0}$ as the initial perturbation. In our calculations, we shall set $\epsilon=10^{-15}$, $\Delta r_*=0.2$, $\Delta t=0.1$ and $\Psi_{j,0}=\exp[-\frac{(r_j-10)^2}{8}]$ and $\Psi_{j,i<0}=0$, and similar settings have also been used in \cite{Zhu:2014sya, Yang:2021yoe, Fu:2022cul} and references therein.

\end{appendices}

\bibliography{ref}

\begin{thebibliography}{81}
\expandafter\ifx\csname natexlab\endcsname\relax\def\natexlab#1{#1}\fi
\expandafter\ifx\csname bibnamefont\endcsname\relax
  \def\bibnamefont#1{#1}\fi
\expandafter\ifx\csname bibfnamefont\endcsname\relax
  \def\bibfnamefont#1{#1}\fi
\expandafter\ifx\csname citenamefont\endcsname\relax
  \def\citenamefont#1{#1}\fi
\expandafter\ifx\csname url\endcsname\relax
  \def\url#1{\texttt{#1}}\fi
\expandafter\ifx\csname urlprefix\endcsname\relax\def\urlprefix{URL }\fi
\providecommand{\bibinfo}[2]{#2}
\providecommand{\eprint}[2][]{\url{#2}}

\bibitem[{\citenamefont{Abbott et~al.}(2016)}]{LIGOScientific:2016aoc}
\bibinfo{author}{\bibfnamefont{B.~P.} \bibnamefont{Abbott}}
  \bibnamefont{et~al.} (\bibinfo{collaboration}{LIGO Scientific, Virgo}),
  \bibinfo{journal}{Phys. Rev. Lett.} \textbf{\bibinfo{volume}{116}},
  \bibinfo{pages}{061102} (\bibinfo{year}{2016}), \eprint{1602.03837}.

\bibitem[{\citenamefont{Abbott et~al.}(2019)}]{LIGOScientific:2018mvr}
\bibinfo{author}{\bibfnamefont{B.~P.} \bibnamefont{Abbott}}
  \bibnamefont{et~al.} (\bibinfo{collaboration}{LIGO Scientific, Virgo}),
  \bibinfo{journal}{Phys. Rev. X} \textbf{\bibinfo{volume}{9}},
  \bibinfo{pages}{031040} (\bibinfo{year}{2019}), \eprint{1811.12907}.

\bibitem[{\citenamefont{Abbott et~al.}(2020)}]{LIGOScientific:2020aai}
\bibinfo{author}{\bibfnamefont{B.~P.} \bibnamefont{Abbott}}
  \bibnamefont{et~al.} (\bibinfo{collaboration}{LIGO Scientific, Virgo}),
  \bibinfo{journal}{Astrophys. J. Lett.} \textbf{\bibinfo{volume}{892}},
  \bibinfo{pages}{L3} (\bibinfo{year}{2020}), \eprint{2001.01761}.

\bibitem[{\citenamefont{Akiyama et~al.}(2019)}]{EventHorizonTelescope:2019dse}
\bibinfo{author}{\bibfnamefont{K.}~\bibnamefont{Akiyama}} \bibnamefont{et~al.}
  (\bibinfo{collaboration}{Event Horizon Telescope}),
  \bibinfo{journal}{Astrophys. J. Lett.} \textbf{\bibinfo{volume}{875}},
  \bibinfo{pages}{L1} (\bibinfo{year}{2019}), \eprint{1906.11238}.

\bibitem[{\citenamefont{Akiyama et~al.}(2022)}]{EventHorizonTelescope:2022xnr}
\bibinfo{author}{\bibfnamefont{K.}~\bibnamefont{Akiyama}} \bibnamefont{et~al.}
  (\bibinfo{collaboration}{Event Horizon Telescope}),
  \bibinfo{journal}{Astrophys. J. Lett.} \textbf{\bibinfo{volume}{930}},
  \bibinfo{pages}{L12} (\bibinfo{year}{2022}).

\bibitem[{\citenamefont{Nojiri and Odintsov}(2006)}]{Nojiri:2006ri}
\bibinfo{author}{\bibfnamefont{S.}~\bibnamefont{Nojiri}} \bibnamefont{and}
  \bibinfo{author}{\bibfnamefont{S.~D.} \bibnamefont{Odintsov}},
  \bibinfo{journal}{eConf} \textbf{\bibinfo{volume}{C0602061}},
  \bibinfo{pages}{06} (\bibinfo{year}{2006}), \eprint{hep-th/0601213}.

\bibitem[{\citenamefont{Clifton et~al.}(2012)\citenamefont{Clifton, Ferreira,
  Padilla, and Skordis}}]{Clifton:2011jh}
\bibinfo{author}{\bibfnamefont{T.}~\bibnamefont{Clifton}},
  \bibinfo{author}{\bibfnamefont{P.~G.} \bibnamefont{Ferreira}},
  \bibinfo{author}{\bibfnamefont{A.}~\bibnamefont{Padilla}}, \bibnamefont{and}
  \bibinfo{author}{\bibfnamefont{C.}~\bibnamefont{Skordis}},
  \bibinfo{journal}{Phys. Rept.} \textbf{\bibinfo{volume}{513}},
  \bibinfo{pages}{1} (\bibinfo{year}{2012}), \eprint{1106.2476}.

\bibitem[{\citenamefont{Berti et~al.}(2015)}]{Berti:2015itd}
\bibinfo{author}{\bibfnamefont{E.}~\bibnamefont{Berti}} \bibnamefont{et~al.},
  \bibinfo{journal}{Class. Quant. Grav.} \textbf{\bibinfo{volume}{32}},
  \bibinfo{pages}{243001} (\bibinfo{year}{2015}), \eprint{1501.07274}.

\bibitem[{\citenamefont{Damour and Esposito-Farese}(1992)}]{Damour:1992we}
\bibinfo{author}{\bibfnamefont{T.}~\bibnamefont{Damour}} \bibnamefont{and}
  \bibinfo{author}{\bibfnamefont{G.}~\bibnamefont{Esposito-Farese}},
  \bibinfo{journal}{Class. Quant. Grav.} \textbf{\bibinfo{volume}{9}},
  \bibinfo{pages}{2093} (\bibinfo{year}{1992}).

\bibitem[{\citenamefont{Horndeski}(1974)}]{Horndeski:1974wa}
\bibinfo{author}{\bibfnamefont{G.~W.} \bibnamefont{Horndeski}},
  \bibinfo{journal}{Int. J. Theor. Phys.} \textbf{\bibinfo{volume}{10}},
  \bibinfo{pages}{363} (\bibinfo{year}{1974}).

\bibitem[{\citenamefont{Bellini et~al.}(2016)\citenamefont{Bellini, Cuesta,
  Jimenez, and Verde}}]{Bellini:2015xja}
\bibinfo{author}{\bibfnamefont{E.}~\bibnamefont{Bellini}},
  \bibinfo{author}{\bibfnamefont{A.~J.} \bibnamefont{Cuesta}},
  \bibinfo{author}{\bibfnamefont{R.}~\bibnamefont{Jimenez}}, \bibnamefont{and}
  \bibinfo{author}{\bibfnamefont{L.}~\bibnamefont{Verde}},
  \bibinfo{journal}{JCAP} \textbf{\bibinfo{volume}{02}}, \bibinfo{pages}{053}
  (\bibinfo{year}{2016}), \bibinfo{note}{[Erratum: JCAP 06, E01 (2016)]},
  \eprint{1509.07816}.

\bibitem[{\citenamefont{Bhattacharya and
  Chakraborty}(2017)}]{Bhattacharya:2016naa}
\bibinfo{author}{\bibfnamefont{S.}~\bibnamefont{Bhattacharya}}
  \bibnamefont{and}
  \bibinfo{author}{\bibfnamefont{S.}~\bibnamefont{Chakraborty}},
  \bibinfo{journal}{Phys. Rev. D} \textbf{\bibinfo{volume}{95}},
  \bibinfo{pages}{044037} (\bibinfo{year}{2017}), \eprint{1607.03693}.

\bibitem[{\citenamefont{Kreisch and Komatsu}(2018)}]{Kreisch:2017uet}
\bibinfo{author}{\bibfnamefont{C.~D.} \bibnamefont{Kreisch}} \bibnamefont{and}
  \bibinfo{author}{\bibfnamefont{E.}~\bibnamefont{Komatsu}},
  \bibinfo{journal}{JCAP} \textbf{\bibinfo{volume}{12}}, \bibinfo{pages}{030}
  (\bibinfo{year}{2018}), \eprint{1712.02710}.

\bibitem[{\citenamefont{Hou and Gong}(2018)}]{Hou:2017cjy}
\bibinfo{author}{\bibfnamefont{S.}~\bibnamefont{Hou}} \bibnamefont{and}
  \bibinfo{author}{\bibfnamefont{Y.}~\bibnamefont{Gong}},
  \bibinfo{journal}{Eur. Phys. J. C} \textbf{\bibinfo{volume}{78}},
  \bibinfo{pages}{247} (\bibinfo{year}{2018}), \eprint{1711.05034}.

\bibitem[{\citenamefont{Spurio~Mancini
  et~al.}(2019)\citenamefont{Spurio~Mancini, K\"ohlinger, Joachimi, Pettorino,
  Sch\"afer, Reischke, van Uitert, Brieden, Archidiacono, and
  Lesgourgues}}]{SpurioMancini:2019rxy}
\bibinfo{author}{\bibfnamefont{A.}~\bibnamefont{Spurio~Mancini}},
  \bibinfo{author}{\bibfnamefont{F.}~\bibnamefont{K\"ohlinger}},
  \bibinfo{author}{\bibfnamefont{B.}~\bibnamefont{Joachimi}},
  \bibinfo{author}{\bibfnamefont{V.}~\bibnamefont{Pettorino}},
  \bibinfo{author}{\bibfnamefont{B.~M.} \bibnamefont{Sch\"afer}},
  \bibinfo{author}{\bibfnamefont{R.}~\bibnamefont{Reischke}},
  \bibinfo{author}{\bibfnamefont{E.}~\bibnamefont{van Uitert}},
  \bibinfo{author}{\bibfnamefont{S.}~\bibnamefont{Brieden}},
  \bibinfo{author}{\bibfnamefont{M.}~\bibnamefont{Archidiacono}},
  \bibnamefont{and}
  \bibinfo{author}{\bibfnamefont{J.}~\bibnamefont{Lesgourgues}},
  \bibinfo{journal}{Mon. Not. Roy. Astron. Soc.}
  \textbf{\bibinfo{volume}{490}}, \bibinfo{pages}{2155} (\bibinfo{year}{2019}),
  \eprint{1901.03686}.

\bibitem[{\citenamefont{Allahyari et~al.}(2020)\citenamefont{Allahyari, Gorji,
  and Mukohyama}}]{Allahyari:2020jkn}
\bibinfo{author}{\bibfnamefont{A.}~\bibnamefont{Allahyari}},
  \bibinfo{author}{\bibfnamefont{M.~A.} \bibnamefont{Gorji}}, \bibnamefont{and}
  \bibinfo{author}{\bibfnamefont{S.}~\bibnamefont{Mukohyama}},
  \bibinfo{journal}{JCAP} \textbf{\bibinfo{volume}{05}}, \bibinfo{pages}{013}
  (\bibinfo{year}{2020}), \bibinfo{note}{[Erratum: JCAP 05, E02 (2021)]},
  \eprint{2002.11932}.

\bibitem[{\citenamefont{Kobayashi}(2019)}]{Kobayashi:2019hrl}
\bibinfo{author}{\bibfnamefont{T.}~\bibnamefont{Kobayashi}},
  \bibinfo{journal}{Rept. Prog. Phys.} \textbf{\bibinfo{volume}{82}},
  \bibinfo{pages}{086901} (\bibinfo{year}{2019}), \eprint{1901.07183}.

\bibitem[{\citenamefont{Rinaldi}(2012)}]{Rinaldi:2012vy}
\bibinfo{author}{\bibfnamefont{M.}~\bibnamefont{Rinaldi}},
  \bibinfo{journal}{Phys. Rev. D} \textbf{\bibinfo{volume}{86}},
  \bibinfo{pages}{084048} (\bibinfo{year}{2012}), \eprint{1208.0103}.

\bibitem[{\citenamefont{Cisterna and Erices}(2014)}]{Cisterna:2014nua}
\bibinfo{author}{\bibfnamefont{A.}~\bibnamefont{Cisterna}} \bibnamefont{and}
  \bibinfo{author}{\bibfnamefont{C.}~\bibnamefont{Erices}},
  \bibinfo{journal}{Phys. Rev. D} \textbf{\bibinfo{volume}{89}},
  \bibinfo{pages}{084038} (\bibinfo{year}{2014}), \eprint{1401.4479}.

\bibitem[{\citenamefont{Feng et~al.}(2015)\citenamefont{Feng, Liu, L\"u, and
  Pope}}]{Feng:2015oea}
\bibinfo{author}{\bibfnamefont{X.-H.} \bibnamefont{Feng}},
  \bibinfo{author}{\bibfnamefont{H.-S.} \bibnamefont{Liu}},
  \bibinfo{author}{\bibfnamefont{H.}~\bibnamefont{L\"u}}, \bibnamefont{and}
  \bibinfo{author}{\bibfnamefont{C.~N.} \bibnamefont{Pope}},
  \bibinfo{journal}{JHEP} \textbf{\bibinfo{volume}{11}}, \bibinfo{pages}{176}
  (\bibinfo{year}{2015}), \eprint{1509.07142}.

\bibitem[{\citenamefont{Sotiriou and Zhou}(2014)}]{Sotiriou:2013qea}
\bibinfo{author}{\bibfnamefont{T.~P.} \bibnamefont{Sotiriou}} \bibnamefont{and}
  \bibinfo{author}{\bibfnamefont{S.-Y.} \bibnamefont{Zhou}},
  \bibinfo{journal}{Phys. Rev. Lett.} \textbf{\bibinfo{volume}{112}},
  \bibinfo{pages}{251102} (\bibinfo{year}{2014}), \eprint{1312.3622}.

\bibitem[{\citenamefont{Miao and Xu}(2016)}]{Miao:2016aol}
\bibinfo{author}{\bibfnamefont{Y.-G.} \bibnamefont{Miao}} \bibnamefont{and}
  \bibinfo{author}{\bibfnamefont{Z.-M.} \bibnamefont{Xu}},
  \bibinfo{journal}{Eur. Phys. J. C} \textbf{\bibinfo{volume}{76}},
  \bibinfo{pages}{638} (\bibinfo{year}{2016}), \eprint{1607.06629}.

\bibitem[{\citenamefont{Kuang and Papantonopoulos}(2016)}]{Kuang:2016edj}
\bibinfo{author}{\bibfnamefont{X.-M.} \bibnamefont{Kuang}} \bibnamefont{and}
  \bibinfo{author}{\bibfnamefont{E.}~\bibnamefont{Papantonopoulos}},
  \bibinfo{journal}{JHEP} \textbf{\bibinfo{volume}{08}}, \bibinfo{pages}{161}
  (\bibinfo{year}{2016}), \eprint{1607.04928}.

\bibitem[{\citenamefont{Babichev et~al.}(2016)\citenamefont{Babichev,
  Charmousis, and Leh\'ebel}}]{Babichev:2016rlq}
\bibinfo{author}{\bibfnamefont{E.}~\bibnamefont{Babichev}},
  \bibinfo{author}{\bibfnamefont{C.}~\bibnamefont{Charmousis}},
  \bibnamefont{and}
  \bibinfo{author}{\bibfnamefont{A.}~\bibnamefont{Leh\'ebel}},
  \bibinfo{journal}{Class. Quant. Grav.} \textbf{\bibinfo{volume}{33}},
  \bibinfo{pages}{154002} (\bibinfo{year}{2016}), \eprint{1604.06402}.

\bibitem[{\citenamefont{Benkel et~al.}(2017)\citenamefont{Benkel, Sotiriou, and
  Witek}}]{Benkel:2016rlz}
\bibinfo{author}{\bibfnamefont{R.}~\bibnamefont{Benkel}},
  \bibinfo{author}{\bibfnamefont{T.~P.} \bibnamefont{Sotiriou}},
  \bibnamefont{and} \bibinfo{author}{\bibfnamefont{H.}~\bibnamefont{Witek}},
  \bibinfo{journal}{Class. Quant. Grav.} \textbf{\bibinfo{volume}{34}},
  \bibinfo{pages}{064001} (\bibinfo{year}{2017}), \eprint{1610.09168}.

\bibitem[{\citenamefont{Filios et~al.}(2019)\citenamefont{Filios, Gonz\'alez,
  Kuang, Papantonopoulos, and V\'asquez}}]{Filios:2018xvy}
\bibinfo{author}{\bibfnamefont{G.}~\bibnamefont{Filios}},
  \bibinfo{author}{\bibfnamefont{P.~A.} \bibnamefont{Gonz\'alez}},
  \bibinfo{author}{\bibfnamefont{X.-M.} \bibnamefont{Kuang}},
  \bibinfo{author}{\bibfnamefont{E.}~\bibnamefont{Papantonopoulos}},
  \bibnamefont{and}
  \bibinfo{author}{\bibfnamefont{Y.}~\bibnamefont{V\'asquez}},
  \bibinfo{journal}{Phys. Rev. D} \textbf{\bibinfo{volume}{99}},
  \bibinfo{pages}{046017} (\bibinfo{year}{2019}), \eprint{1808.07766}.

\bibitem[{\citenamefont{Cisterna et~al.}(2018)\citenamefont{Cisterna, Erices,
  Kuang, and Rinaldi}}]{Cisterna:2018hzf}
\bibinfo{author}{\bibfnamefont{A.}~\bibnamefont{Cisterna}},
  \bibinfo{author}{\bibfnamefont{C.}~\bibnamefont{Erices}},
  \bibinfo{author}{\bibfnamefont{X.-M.} \bibnamefont{Kuang}}, \bibnamefont{and}
  \bibinfo{author}{\bibfnamefont{M.}~\bibnamefont{Rinaldi}},
  \bibinfo{journal}{Phys. Rev. D} \textbf{\bibinfo{volume}{97}},
  \bibinfo{pages}{124052} (\bibinfo{year}{2018}), \eprint{1803.07600}.

\bibitem[{\citenamefont{Giusti et~al.}(2022)\citenamefont{Giusti, Zentarra,
  Heisenberg, and Faraoni}}]{Giusti:2021sku}
\bibinfo{author}{\bibfnamefont{A.}~\bibnamefont{Giusti}},
  \bibinfo{author}{\bibfnamefont{S.}~\bibnamefont{Zentarra}},
  \bibinfo{author}{\bibfnamefont{L.}~\bibnamefont{Heisenberg}},
  \bibnamefont{and} \bibinfo{author}{\bibfnamefont{V.}~\bibnamefont{Faraoni}},
  \bibinfo{journal}{Phys. Rev. D} \textbf{\bibinfo{volume}{105}},
  \bibinfo{pages}{124011} (\bibinfo{year}{2022}), \eprint{2108.10706}.

\bibitem[{\citenamefont{Babichev and Charmousis}(2014)}]{Babichev:2013cya}
\bibinfo{author}{\bibfnamefont{E.}~\bibnamefont{Babichev}} \bibnamefont{and}
  \bibinfo{author}{\bibfnamefont{C.}~\bibnamefont{Charmousis}},
  \bibinfo{journal}{JHEP} \textbf{\bibinfo{volume}{08}}, \bibinfo{pages}{106}
  (\bibinfo{year}{2014}), \eprint{1312.3204}.

\bibitem[{\citenamefont{Babichev et~al.}(2018)\citenamefont{Babichev,
  Charmousis, Esposito-Far\`ese, and Leh\'ebel}}]{Babichev:2017lmw}
\bibinfo{author}{\bibfnamefont{E.}~\bibnamefont{Babichev}},
  \bibinfo{author}{\bibfnamefont{C.}~\bibnamefont{Charmousis}},
  \bibinfo{author}{\bibfnamefont{G.}~\bibnamefont{Esposito-Far\`ese}},
  \bibnamefont{and}
  \bibinfo{author}{\bibfnamefont{A.}~\bibnamefont{Leh\'ebel}},
  \bibinfo{journal}{Phys. Rev. Lett.} \textbf{\bibinfo{volume}{120}},
  \bibinfo{pages}{241101} (\bibinfo{year}{2018}), \eprint{1712.04398}.

\bibitem[{\citenamefont{Ben~Achour and Liu}(2019)}]{BenAchour:2018dap}
\bibinfo{author}{\bibfnamefont{J.}~\bibnamefont{Ben~Achour}} \bibnamefont{and}
  \bibinfo{author}{\bibfnamefont{H.}~\bibnamefont{Liu}},
  \bibinfo{journal}{Phys. Rev. D} \textbf{\bibinfo{volume}{99}},
  \bibinfo{pages}{064042} (\bibinfo{year}{2019}), \eprint{1811.05369}.

\bibitem[{\citenamefont{Takahashi et~al.}(2019)\citenamefont{Takahashi,
  Motohashi, and Minamitsuji}}]{Takahashi:2019oxz}
\bibinfo{author}{\bibfnamefont{K.}~\bibnamefont{Takahashi}},
  \bibinfo{author}{\bibfnamefont{H.}~\bibnamefont{Motohashi}},
  \bibnamefont{and}
  \bibinfo{author}{\bibfnamefont{M.}~\bibnamefont{Minamitsuji}},
  \bibinfo{journal}{Phys. Rev. D} \textbf{\bibinfo{volume}{100}},
  \bibinfo{pages}{024041} (\bibinfo{year}{2019}), \eprint{1904.03554}.

\bibitem[{\citenamefont{Minamitsuji and Edholm}(2019)}]{Minamitsuji:2019shy}
\bibinfo{author}{\bibfnamefont{M.}~\bibnamefont{Minamitsuji}} \bibnamefont{and}
  \bibinfo{author}{\bibfnamefont{J.}~\bibnamefont{Edholm}},
  \bibinfo{journal}{Phys. Rev. D} \textbf{\bibinfo{volume}{100}},
  \bibinfo{pages}{044053} (\bibinfo{year}{2019}), \eprint{1907.02072}.

\bibitem[{\citenamefont{Arkani-Hamed et~al.}(2004)\citenamefont{Arkani-Hamed,
  Creminelli, Mukohyama, and Zaldarriaga}}]{Arkani-Hamed:2003juy}
\bibinfo{author}{\bibfnamefont{N.}~\bibnamefont{Arkani-Hamed}},
  \bibinfo{author}{\bibfnamefont{P.}~\bibnamefont{Creminelli}},
  \bibinfo{author}{\bibfnamefont{S.}~\bibnamefont{Mukohyama}},
  \bibnamefont{and}
  \bibinfo{author}{\bibfnamefont{M.}~\bibnamefont{Zaldarriaga}},
  \bibinfo{journal}{JCAP} \textbf{\bibinfo{volume}{04}}, \bibinfo{pages}{001}
  (\bibinfo{year}{2004}), \eprint{hep-th/0312100}.

\bibitem[{\citenamefont{Khoury et~al.}(2020)\citenamefont{Khoury, Trodden, and
  Wong}}]{Khoury:2020aya}
\bibinfo{author}{\bibfnamefont{J.}~\bibnamefont{Khoury}},
  \bibinfo{author}{\bibfnamefont{M.}~\bibnamefont{Trodden}}, \bibnamefont{and}
  \bibinfo{author}{\bibfnamefont{S.~S.~C.} \bibnamefont{Wong}},
  \bibinfo{journal}{JCAP} \textbf{\bibinfo{volume}{11}}, \bibinfo{pages}{044}
  (\bibinfo{year}{2020}), \eprint{2007.01320}.

\bibitem[{\citenamefont{Hui and Nicolis}(2013)}]{Hui:2012qt}
\bibinfo{author}{\bibfnamefont{L.}~\bibnamefont{Hui}} \bibnamefont{and}
  \bibinfo{author}{\bibfnamefont{A.}~\bibnamefont{Nicolis}},
  \bibinfo{journal}{Phys. Rev. Lett.} \textbf{\bibinfo{volume}{110}},
  \bibinfo{pages}{241104} (\bibinfo{year}{2013}), \eprint{1202.1296}.

\bibitem[{\citenamefont{Babichev et~al.}(2017)\citenamefont{Babichev,
  Charmousis, and Leh\'ebel}}]{Babichev:2017guv}
\bibinfo{author}{\bibfnamefont{E.}~\bibnamefont{Babichev}},
  \bibinfo{author}{\bibfnamefont{C.}~\bibnamefont{Charmousis}},
  \bibnamefont{and}
  \bibinfo{author}{\bibfnamefont{A.}~\bibnamefont{Leh\'ebel}},
  \bibinfo{journal}{JCAP} \textbf{\bibinfo{volume}{04}}, \bibinfo{pages}{027}
  (\bibinfo{year}{2017}), \eprint{1702.01938}.

\bibitem[{\citenamefont{Bergliaffa et~al.}(2021)\citenamefont{Bergliaffa,
  Maier, and Silvano}}]{Bergliaffa:2021diw}
\bibinfo{author}{\bibfnamefont{S.~E.~P.} \bibnamefont{Bergliaffa}},
  \bibinfo{author}{\bibfnamefont{R.}~\bibnamefont{Maier}}, \bibnamefont{and}
  \bibinfo{author}{\bibfnamefont{N.~d.~O.} \bibnamefont{Silvano}}
  (\bibinfo{year}{2021}), \eprint{2107.07839}.

\bibitem[{\citenamefont{Kumar et~al.}(2022)\citenamefont{Kumar, Islam, and
  Ghosh}}]{Kumar:2021cyl}
\bibinfo{author}{\bibfnamefont{J.}~\bibnamefont{Kumar}},
  \bibinfo{author}{\bibfnamefont{S.~U.} \bibnamefont{Islam}}, \bibnamefont{and}
  \bibinfo{author}{\bibfnamefont{S.~G.} \bibnamefont{Ghosh}},
  \bibinfo{journal}{Eur. Phys. J. C} \textbf{\bibinfo{volume}{82}},
  \bibinfo{pages}{443} (\bibinfo{year}{2022}), \eprint{2109.04450}.

\bibitem[{\citenamefont{Walia et~al.}(2022)\citenamefont{Walia, Maharaj, and
  Ghosh}}]{Walia:2021emv}
\bibinfo{author}{\bibfnamefont{R.~K.} \bibnamefont{Walia}},
  \bibinfo{author}{\bibfnamefont{S.~D.} \bibnamefont{Maharaj}},
  \bibnamefont{and} \bibinfo{author}{\bibfnamefont{S.~G.} \bibnamefont{Ghosh}},
  \bibinfo{journal}{Eur. Phys. J. C} \textbf{\bibinfo{volume}{82}},
  \bibinfo{pages}{547} (\bibinfo{year}{2022}), \eprint{2109.08055}.

\bibitem[{\citenamefont{Atamurotov et~al.}(2022)\citenamefont{Atamurotov,
  Sarikulov, Abdujabbarov, and Ahmedov}}]{Atamurotov:2022slw}
\bibinfo{author}{\bibfnamefont{F.}~\bibnamefont{Atamurotov}},
  \bibinfo{author}{\bibfnamefont{F.}~\bibnamefont{Sarikulov}},
  \bibinfo{author}{\bibfnamefont{A.}~\bibnamefont{Abdujabbarov}},
  \bibnamefont{and} \bibinfo{author}{\bibfnamefont{B.}~\bibnamefont{Ahmedov}},
  \bibinfo{journal}{Eur. Phys. J. Plus} \textbf{\bibinfo{volume}{137}},
  \bibinfo{pages}{336} (\bibinfo{year}{2022}).

\bibitem[{\citenamefont{Afrin and Ghosh}(2022)}]{Afrin:2021wlj}
\bibinfo{author}{\bibfnamefont{M.}~\bibnamefont{Afrin}} \bibnamefont{and}
  \bibinfo{author}{\bibfnamefont{S.~G.} \bibnamefont{Ghosh}},
  \bibinfo{journal}{Astrophys. J.} \textbf{\bibinfo{volume}{932}},
  \bibinfo{pages}{51} (\bibinfo{year}{2022}), \eprint{2110.05258}.

\bibitem[{\citenamefont{Jha et~al.}(2022)\citenamefont{Jha, Khodadi, Rahaman,
  and Sheykhi}}]{Jha:2022tdl}
\bibinfo{author}{\bibfnamefont{S.~K.} \bibnamefont{Jha}},
  \bibinfo{author}{\bibfnamefont{M.}~\bibnamefont{Khodadi}},
  \bibinfo{author}{\bibfnamefont{A.}~\bibnamefont{Rahaman}}, \bibnamefont{and}
  \bibinfo{author}{\bibfnamefont{A.}~\bibnamefont{Sheykhi}}
  (\bibinfo{year}{2022}), \eprint{2212.13051}.

\bibitem[{\citenamefont{Wang et~al.}(2023)\citenamefont{Wang, Kuang, Meng,
  Wang, and Wu}}]{Wang:2023vcv}
\bibinfo{author}{\bibfnamefont{X.-J.} \bibnamefont{Wang}},
  \bibinfo{author}{\bibfnamefont{X.-M.} \bibnamefont{Kuang}},
  \bibinfo{author}{\bibfnamefont{Y.}~\bibnamefont{Meng}},
  \bibinfo{author}{\bibfnamefont{B.}~\bibnamefont{Wang}}, \bibnamefont{and}
  \bibinfo{author}{\bibfnamefont{J.-P.} \bibnamefont{Wu}},
  \bibinfo{journal}{Phys. Rev. D} \textbf{\bibinfo{volume}{107}},
  \bibinfo{pages}{124052} (\bibinfo{year}{2023}), \eprint{2304.10015}.

\bibitem[{\citenamefont{Nollert}(1999)}]{Nollert:1999ji}
\bibinfo{author}{\bibfnamefont{H.-P.} \bibnamefont{Nollert}},
  \bibinfo{journal}{Class. Quant. Grav.} \textbf{\bibinfo{volume}{16}},
  \bibinfo{pages}{R159} (\bibinfo{year}{1999}).

\bibitem[{\citenamefont{Berti et~al.}(2009)\citenamefont{Berti, Cardoso, and
  Starinets}}]{Berti:2009kk}
\bibinfo{author}{\bibfnamefont{E.}~\bibnamefont{Berti}},
  \bibinfo{author}{\bibfnamefont{V.}~\bibnamefont{Cardoso}}, \bibnamefont{and}
  \bibinfo{author}{\bibfnamefont{A.~O.} \bibnamefont{Starinets}},
  \bibinfo{journal}{Class. Quant. Grav.} \textbf{\bibinfo{volume}{26}},
  \bibinfo{pages}{163001} (\bibinfo{year}{2009}), \eprint{0905.2975}.

\bibitem[{\citenamefont{Konoplya and Zhidenko}(2011)}]{Konoplya:2011qq}
\bibinfo{author}{\bibfnamefont{R.~A.} \bibnamefont{Konoplya}} \bibnamefont{and}
  \bibinfo{author}{\bibfnamefont{A.}~\bibnamefont{Zhidenko}},
  \bibinfo{journal}{Rev. Mod. Phys.} \textbf{\bibinfo{volume}{83}},
  \bibinfo{pages}{793} (\bibinfo{year}{2011}), \eprint{1102.4014}.

\bibitem[{\citenamefont{Harmark et~al.}(2010)\citenamefont{Harmark, Natario,
  and Schiappa}}]{Harmark:2007jy}
\bibinfo{author}{\bibfnamefont{T.}~\bibnamefont{Harmark}},
  \bibinfo{author}{\bibfnamefont{J.}~\bibnamefont{Natario}}, \bibnamefont{and}
  \bibinfo{author}{\bibfnamefont{R.}~\bibnamefont{Schiappa}},
  \bibinfo{journal}{Adv. Theor. Math. Phys.} \textbf{\bibinfo{volume}{14}},
  \bibinfo{pages}{727} (\bibinfo{year}{2010}), \eprint{0708.0017}.

\bibitem[{\citenamefont{Kanti and March-Russell}(2002)}]{Kanti:2002nr}
\bibinfo{author}{\bibfnamefont{P.}~\bibnamefont{Kanti}} \bibnamefont{and}
  \bibinfo{author}{\bibfnamefont{J.}~\bibnamefont{March-Russell}},
  \bibinfo{journal}{Phys. Rev. D} \textbf{\bibinfo{volume}{66}},
  \bibinfo{pages}{024023} (\bibinfo{year}{2002}), \eprint{hep-ph/0203223}.

\bibitem[{\citenamefont{Hawking}(1975)}]{Hawking:1975vcx}
\bibinfo{author}{\bibfnamefont{S.~W.} \bibnamefont{Hawking}},
  \bibinfo{journal}{Commun. Math. Phys.} \textbf{\bibinfo{volume}{43}},
  \bibinfo{pages}{199} (\bibinfo{year}{1975}), \bibinfo{note}{[Erratum:
  Commun.Math.Phys. 46, 206 (1976)]}.

\bibitem[{\citenamefont{Regge and Wheeler}(1957)}]{Regge:1957td}
\bibinfo{author}{\bibfnamefont{T.}~\bibnamefont{Regge}} \bibnamefont{and}
  \bibinfo{author}{\bibfnamefont{J.~A.} \bibnamefont{Wheeler}},
  \bibinfo{journal}{Phys. Rev.} \textbf{\bibinfo{volume}{108}},
  \bibinfo{pages}{1063} (\bibinfo{year}{1957}).

\bibitem[{\citenamefont{Zerilli}(1970)}]{Zerilli:1970se}
\bibinfo{author}{\bibfnamefont{F.~J.} \bibnamefont{Zerilli}},
  \bibinfo{journal}{Phys. Rev. Lett.} \textbf{\bibinfo{volume}{24}},
  \bibinfo{pages}{737} (\bibinfo{year}{1970}).

\bibitem[{\citenamefont{Cho}(2003)}]{Cho:2003qe}
\bibinfo{author}{\bibfnamefont{H.~T.} \bibnamefont{Cho}},
  \bibinfo{journal}{Phys. Rev. D} \textbf{\bibinfo{volume}{68}},
  \bibinfo{pages}{024003} (\bibinfo{year}{2003}), \eprint{gr-qc/0303078}.

\bibitem[{\citenamefont{Anderson and Price}(1991)}]{Anderson:1991kx}
\bibinfo{author}{\bibfnamefont{A.}~\bibnamefont{Anderson}} \bibnamefont{and}
  \bibinfo{author}{\bibfnamefont{R.~H.} \bibnamefont{Price}},
  \bibinfo{journal}{Phys. Rev. D} \textbf{\bibinfo{volume}{43}},
  \bibinfo{pages}{3147} (\bibinfo{year}{1991}).

\bibitem[{\citenamefont{Kokkotas and Schmidt}(1999)}]{Kokkotas:1999bd}
\bibinfo{author}{\bibfnamefont{K.~D.} \bibnamefont{Kokkotas}} \bibnamefont{and}
  \bibinfo{author}{\bibfnamefont{B.~G.} \bibnamefont{Schmidt}},
  \bibinfo{journal}{Living Rev. Rel.} \textbf{\bibinfo{volume}{2}},
  \bibinfo{pages}{2} (\bibinfo{year}{1999}), \eprint{gr-qc/9909058}.

\bibitem[{\citenamefont{Matyjasek and Opala}(2017)}]{Matyjasek:2017psv}
\bibinfo{author}{\bibfnamefont{J.}~\bibnamefont{Matyjasek}} \bibnamefont{and}
  \bibinfo{author}{\bibfnamefont{M.}~\bibnamefont{Opala}},
  \bibinfo{journal}{Phys. Rev. D} \textbf{\bibinfo{volume}{96}},
  \bibinfo{pages}{024011} (\bibinfo{year}{2017}), \eprint{1704.00361}.

\bibitem[{\citenamefont{Konoplya et~al.}(2019)\citenamefont{Konoplya, Zhidenko,
  and Zinhailo}}]{Konoplya:2019hlu}
\bibinfo{author}{\bibfnamefont{R.~A.} \bibnamefont{Konoplya}},
  \bibinfo{author}{\bibfnamefont{A.}~\bibnamefont{Zhidenko}}, \bibnamefont{and}
  \bibinfo{author}{\bibfnamefont{A.~F.} \bibnamefont{Zinhailo}},
  \bibinfo{journal}{Class. Quant. Grav.} \textbf{\bibinfo{volume}{36}},
  \bibinfo{pages}{155002} (\bibinfo{year}{2019}), \eprint{1904.10333}.

\bibitem[{\citenamefont{Zhang et~al.}(2007)\citenamefont{Zhang, Gui, and
  Li}}]{Zhang:2006hh}
\bibinfo{author}{\bibfnamefont{Y.}~\bibnamefont{Zhang}},
  \bibinfo{author}{\bibfnamefont{Y.~X.} \bibnamefont{Gui}}, \bibnamefont{and}
  \bibinfo{author}{\bibfnamefont{F.}~\bibnamefont{Li}}, \bibinfo{journal}{Gen.
  Rel. Grav.} \textbf{\bibinfo{volume}{39}}, \bibinfo{pages}{1003}
  (\bibinfo{year}{2007}), \eprint{gr-qc/0612010}.

\bibitem[{\citenamefont{Schutz and Will}(1985)}]{Schutz:1985km}
\bibinfo{author}{\bibfnamefont{B.~F.} \bibnamefont{Schutz}} \bibnamefont{and}
  \bibinfo{author}{\bibfnamefont{C.~M.} \bibnamefont{Will}},
  \bibinfo{journal}{Astrophys. J. Lett.} \textbf{\bibinfo{volume}{291}},
  \bibinfo{pages}{L33} (\bibinfo{year}{1985}).

\bibitem[{\citenamefont{Cardoso et~al.}(2009)\citenamefont{Cardoso, Miranda,
  Berti, Witek, and Zanchin}}]{Cardoso:2008bp}
\bibinfo{author}{\bibfnamefont{V.}~\bibnamefont{Cardoso}},
  \bibinfo{author}{\bibfnamefont{A.~S.} \bibnamefont{Miranda}},
  \bibinfo{author}{\bibfnamefont{E.}~\bibnamefont{Berti}},
  \bibinfo{author}{\bibfnamefont{H.}~\bibnamefont{Witek}}, \bibnamefont{and}
  \bibinfo{author}{\bibfnamefont{V.~T.} \bibnamefont{Zanchin}},
  \bibinfo{journal}{Phys. Rev. D} \textbf{\bibinfo{volume}{79}},
  \bibinfo{pages}{064016} (\bibinfo{year}{2009}), \eprint{0812.1806}.

\bibitem[{\citenamefont{Jusufi}(2020{\natexlab{a}})}]{Jusufi:2019ltj}
\bibinfo{author}{\bibfnamefont{K.}~\bibnamefont{Jusufi}},
  \bibinfo{journal}{Phys. Rev. D} \textbf{\bibinfo{volume}{101}},
  \bibinfo{pages}{084055} (\bibinfo{year}{2020}{\natexlab{a}}),
  \eprint{1912.13320}.

\bibitem[{\citenamefont{Jusufi}(2020{\natexlab{b}})}]{Jusufi:2020dhz}
\bibinfo{author}{\bibfnamefont{K.}~\bibnamefont{Jusufi}},
  \bibinfo{journal}{Phys. Rev. D} \textbf{\bibinfo{volume}{101}},
  \bibinfo{pages}{124063} (\bibinfo{year}{2020}{\natexlab{b}}),
  \eprint{2004.04664}.

\bibitem[{\citenamefont{Teukolsky}(1973)}]{Teukolsky:1973ha}
\bibinfo{author}{\bibfnamefont{S.~A.} \bibnamefont{Teukolsky}},
  \bibinfo{journal}{Astrophys. J.} \textbf{\bibinfo{volume}{185}},
  \bibinfo{pages}{635} (\bibinfo{year}{1973}).

\bibitem[{\citenamefont{Jing}(2005)}]{Jing:2005ux}
\bibinfo{author}{\bibfnamefont{J.-l.} \bibnamefont{Jing}}
  (\bibinfo{year}{2005}), \eprint{gr-qc/0502010}.

\bibitem[{\citenamefont{Harris and Kanti}(2003)}]{Harris:2003eg}
\bibinfo{author}{\bibfnamefont{C.~M.} \bibnamefont{Harris}} \bibnamefont{and}
  \bibinfo{author}{\bibfnamefont{P.}~\bibnamefont{Kanti}},
  \bibinfo{journal}{JHEP} \textbf{\bibinfo{volume}{10}}, \bibinfo{pages}{014}
  (\bibinfo{year}{2003}), \eprint{hep-ph/0309054}.

\bibitem[{\citenamefont{Chandrasekhar}(1975)}]{Chandrasekhar:1975nkd}
\bibinfo{author}{\bibfnamefont{S.}~\bibnamefont{Chandrasekhar}},
  \bibinfo{journal}{Proc. Roy. Soc. Lond. A} \textbf{\bibinfo{volume}{343}},
  \bibinfo{pages}{289} (\bibinfo{year}{1975}).

\bibitem[{\citenamefont{Iyer and Will}(1987)}]{PhysRevD.35.3621}
\bibinfo{author}{\bibfnamefont{S.}~\bibnamefont{Iyer}} \bibnamefont{and}
  \bibinfo{author}{\bibfnamefont{C.~M.} \bibnamefont{Will}},
  \bibinfo{journal}{Phys. Rev. D} \textbf{\bibinfo{volume}{35}},
  \bibinfo{pages}{3621} (\bibinfo{year}{1987}).

\bibitem[{\citenamefont{Page}(1976)}]{PhysRevD.13.198}
\bibinfo{author}{\bibfnamefont{D.~N.} \bibnamefont{Page}},
  \bibinfo{journal}{Phys. Rev. D} \textbf{\bibinfo{volume}{13}},
  \bibinfo{pages}{198} (\bibinfo{year}{1976}).

\bibitem[{\citenamefont{Adams}(2000)}]{Adams:2000ax}
\bibinfo{author}{\bibfnamefont{F.~C.} \bibnamefont{Adams}},
  \bibinfo{journal}{Gen. Rel. Grav.} \textbf{\bibinfo{volume}{32}},
  \bibinfo{pages}{2229} (\bibinfo{year}{2000}), \eprint{gr-qc/0006062}.

\bibitem[{\citenamefont{Giesler et~al.}(2019)\citenamefont{Giesler, Isi,
  Scheel, and Teukolsky}}]{Giesler:2019uxc}
\bibinfo{author}{\bibfnamefont{M.}~\bibnamefont{Giesler}},
  \bibinfo{author}{\bibfnamefont{M.}~\bibnamefont{Isi}},
  \bibinfo{author}{\bibfnamefont{M.~A.} \bibnamefont{Scheel}},
  \bibnamefont{and}
  \bibinfo{author}{\bibfnamefont{S.}~\bibnamefont{Teukolsky}},
  \bibinfo{journal}{Phys. Rev. X} \textbf{\bibinfo{volume}{9}},
  \bibinfo{pages}{041060} (\bibinfo{year}{2019}), \eprint{1903.08284}.

\bibitem[{\citenamefont{Sago et~al.}(2021)\citenamefont{Sago, Isoyama, and
  Nakano}}]{Sago:2021gbq}
\bibinfo{author}{\bibfnamefont{N.}~\bibnamefont{Sago}},
  \bibinfo{author}{\bibfnamefont{S.}~\bibnamefont{Isoyama}}, \bibnamefont{and}
  \bibinfo{author}{\bibfnamefont{H.}~\bibnamefont{Nakano}},
  \bibinfo{journal}{Universe} \textbf{\bibinfo{volume}{7}},
  \bibinfo{pages}{357} (\bibinfo{year}{2021}), \eprint{2108.13017}.

\bibitem[{\citenamefont{Konoplya and Zhidenko}(2005)}]{Konoplya:2004wg}
\bibinfo{author}{\bibfnamefont{R.~A.} \bibnamefont{Konoplya}} \bibnamefont{and}
  \bibinfo{author}{\bibfnamefont{A.~V.} \bibnamefont{Zhidenko}},
  \bibinfo{journal}{Phys. Lett. B} \textbf{\bibinfo{volume}{609}},
  \bibinfo{pages}{377} (\bibinfo{year}{2005}), \eprint{gr-qc/0411059}.

\bibitem[{\citenamefont{Tattersall and Ferreira}(2018)}]{Tattersall:2018nve}
\bibinfo{author}{\bibfnamefont{O.~J.} \bibnamefont{Tattersall}}
  \bibnamefont{and} \bibinfo{author}{\bibfnamefont{P.~G.}
  \bibnamefont{Ferreira}}, \bibinfo{journal}{Phys. Rev. D}
  \textbf{\bibinfo{volume}{97}}, \bibinfo{pages}{104047}
  (\bibinfo{year}{2018}), \eprint{1804.08950}.

\bibitem[{\citenamefont{Konoplya}(2003)}]{Konoplya:2003ii}
\bibinfo{author}{\bibfnamefont{R.~A.} \bibnamefont{Konoplya}},
  \bibinfo{journal}{Phys. Rev. D} \textbf{\bibinfo{volume}{68}},
  \bibinfo{pages}{024018} (\bibinfo{year}{2003}), \eprint{gr-qc/0303052}.

\bibitem[{\citenamefont{Lin and Qian}(2017)}]{Lin:2016sch}
\bibinfo{author}{\bibfnamefont{K.}~\bibnamefont{Lin}} \bibnamefont{and}
  \bibinfo{author}{\bibfnamefont{W.-L.} \bibnamefont{Qian}},
  \bibinfo{journal}{Class. Quant. Grav.} \textbf{\bibinfo{volume}{34}},
  \bibinfo{pages}{095004} (\bibinfo{year}{2017}), \eprint{1610.08135}.

\bibitem[{\citenamefont{Lin et~al.}(2017)\citenamefont{Lin, Qian, Pavan, and
  Abdalla}}]{Lin:2017oag}
\bibinfo{author}{\bibfnamefont{K.}~\bibnamefont{Lin}},
  \bibinfo{author}{\bibfnamefont{W.-L.} \bibnamefont{Qian}},
  \bibinfo{author}{\bibfnamefont{A.~B.} \bibnamefont{Pavan}}, \bibnamefont{and}
  \bibinfo{author}{\bibfnamefont{E.}~\bibnamefont{Abdalla}},
  \bibinfo{journal}{Mod. Phys. Lett. A} \textbf{\bibinfo{volume}{32}},
  \bibinfo{pages}{1750134} (\bibinfo{year}{2017}), \eprint{1703.06439}.

\bibitem[{\citenamefont{Lin and Qian}(2019)}]{Lin:2019mmf}
\bibinfo{author}{\bibfnamefont{K.}~\bibnamefont{Lin}} \bibnamefont{and}
  \bibinfo{author}{\bibfnamefont{W.-L.} \bibnamefont{Qian}},
  \bibinfo{journal}{Chin. Phys. C} \textbf{\bibinfo{volume}{43}},
  \bibinfo{pages}{035105} (\bibinfo{year}{2019}), \eprint{1902.08352}.

\bibitem[{\citenamefont{Abdalla et~al.}(2010)\citenamefont{Abdalla, Pellicer,
  de~Oliveira, and Pavan}}]{Abdalla:2010nq}
\bibinfo{author}{\bibfnamefont{E.}~\bibnamefont{Abdalla}},
  \bibinfo{author}{\bibfnamefont{C.~E.} \bibnamefont{Pellicer}},
  \bibinfo{author}{\bibfnamefont{J.}~\bibnamefont{de~Oliveira}},
  \bibnamefont{and} \bibinfo{author}{\bibfnamefont{A.~B.} \bibnamefont{Pavan}},
  \bibinfo{journal}{Phys. Rev. D} \textbf{\bibinfo{volume}{82}},
  \bibinfo{pages}{124033} (\bibinfo{year}{2010}), \eprint{1010.2806}.

\bibitem[{\citenamefont{Zhu et~al.}(2014)\citenamefont{Zhu, Zhang, Pellicer,
  Wang, and Abdalla}}]{Zhu:2014sya}
\bibinfo{author}{\bibfnamefont{Z.}~\bibnamefont{Zhu}},
  \bibinfo{author}{\bibfnamefont{S.-J.} \bibnamefont{Zhang}},
  \bibinfo{author}{\bibfnamefont{C.~E.} \bibnamefont{Pellicer}},
  \bibinfo{author}{\bibfnamefont{B.}~\bibnamefont{Wang}}, \bibnamefont{and}
  \bibinfo{author}{\bibfnamefont{E.}~\bibnamefont{Abdalla}},
  \bibinfo{journal}{Phys. Rev. D} \textbf{\bibinfo{volume}{90}},
  \bibinfo{pages}{044042} (\bibinfo{year}{2014}), \bibinfo{note}{[Addendum:
  Phys.Rev.D 90, 049904 (2014)]}, \eprint{1405.4931}.

\bibitem[{\citenamefont{Yang et~al.}(2022)\citenamefont{Yang, Fu, Kuang, and
  Wu}}]{Yang:2021yoe}
\bibinfo{author}{\bibfnamefont{Z.-H.} \bibnamefont{Yang}},
  \bibinfo{author}{\bibfnamefont{G.}~\bibnamefont{Fu}},
  \bibinfo{author}{\bibfnamefont{X.-M.} \bibnamefont{Kuang}}, \bibnamefont{and}
  \bibinfo{author}{\bibfnamefont{J.-P.} \bibnamefont{Wu}},
  \bibinfo{journal}{Eur. Phys. J. C} \textbf{\bibinfo{volume}{82}},
  \bibinfo{pages}{868} (\bibinfo{year}{2022}), \eprint{2112.15052}.

\bibitem[{\citenamefont{Fu et~al.}(2023)\citenamefont{Fu, Zhang, Liu, Kuang,
  Pan, and Wu}}]{Fu:2022cul}
\bibinfo{author}{\bibfnamefont{G.}~\bibnamefont{Fu}},
  \bibinfo{author}{\bibfnamefont{D.}~\bibnamefont{Zhang}},
  \bibinfo{author}{\bibfnamefont{P.}~\bibnamefont{Liu}},
  \bibinfo{author}{\bibfnamefont{X.-M.} \bibnamefont{Kuang}},
  \bibinfo{author}{\bibfnamefont{Q.}~\bibnamefont{Pan}}, \bibnamefont{and}
  \bibinfo{author}{\bibfnamefont{J.-P.} \bibnamefont{Wu}},
  \bibinfo{journal}{Phys. Rev. D} \textbf{\bibinfo{volume}{107}},
  \bibinfo{pages}{044049} (\bibinfo{year}{2023}), \eprint{2207.12927}.

\end{thebibliography}
\bibliographystyle{apsrev}

\end{document}